\pdfoutput=1
\documentclass[aip,pof,amsmath,amssymb,reprint]{revtex4-2}
\usepackage{booktabs}
\usepackage{graphicx}
\usepackage{bm}
\usepackage{url}

\begin{document}

\title{A closed-form solution for streaming and Lagrangian transport in a deforming circular cavity}

\author{Zijian~Liu}
\thanks{Z. Liu and J. Olszewski contributed equally to this work.}
\affiliation{Department of Mechanical and Industrial Engineering,
University of Illinois Chicago, Chicago, Illinois 60607, USA}

\author{Julian~Olszewski}
\affiliation{Department of Mechanical and Industrial Engineering,
University of Illinois Chicago, Chicago, Illinois 60607, USA}

\author{Yang~Lin}
\affiliation{Department of Mechanical, Industrial and Systems Engineering,
University of Rhode Island, Kingston, Rhode Island 02881, USA}

\author{Yuan~Gao}
\affiliation{Department of Mechanical Engineering,
University of Memphis, Memphis, Tennessee 38152, USA}

\author{Mengren~Wu}
\affiliation{Department of Mechanical Engineering,
Stanford University, Stanford, California 94305, USA}

\author{Jie~Xu}
\email[Author to whom correspondence should be addressed: ]{jiexu@uic.edu}
\affiliation{Department of Mechanical and Industrial Engineering,
University of Illinois Chicago, Chicago, Illinois 60607, USA}

\date{\today}

\begin{abstract}
Streaming from a deforming cavity wall serves micromixing, pumping and particle handling.
We solve it in closed form in a two-dimensional circular cavity, for any azimuthal
wall mode $m$, as $\mathrm{Wo}^2 \to 0$. A biharmonic inversion against the Reynolds stress,
corrected by the second-order slip a moving wall imposes, gives the Lagrangian mean a tracer
follows for a deforming no-slip wall,
$\psi_L = -[m(5m+4)a_m^2/(128(m+2)(2m+1))]\,r^{2m}(r^2-1)^2\sin 2m\theta$, with a companion form for a
shear-free interface. For a single mode the factor $(5m+4)/(m+2)$ relating it to the auxiliary
solution $\psi_2$ is the same for every member of the co-phased prescribed-velocity family; at
$m=2$ the physical Eulerian
mean peaks an order of magnitude above $\psi_2$, with opposite sign. The no-slip cell centers lie
at $r^2 = m/(m+2)$, and at large $m$ the peak streamfunction falls as $m^{-2}$, the peak speed
as $m^{-1}$. At fixed radial wall-velocity amplitude the ranking over $m$ follows the wall
kinematics: an externally driven wall peaks at $m=1$, a wall with zero first-order surface strain
at $m=3$. Mode superpositions invert without degenerating, each harmonic carrying its own
correction. At finite $\mathrm{Wo}$ the first-order field stays closed form in Bessel functions and
the mean flow reduces to quadrature; the construction approaches the $m=2$ Rayleigh limit on a
separate tangentially driven boundary problem. An independent finite-element solver, with the closed form
withheld, reproduces $\psi_2$ with quadratic mesh convergence.
\end{abstract}

\maketitle

\section{Introduction}
\label{sec:intro}

Acoustically driven microfluidics is now a broad toolkit for manipulating fluid and suspended matter~\cite{friend2011,ozcelik2018}, and bubbles trapped in microcavities are among its more
versatile actuators~\cite{hashmi2012,gao2020review}. Tovar and Lee established the side-cavity
geometry as a lateral cavity acoustic transducer~\cite{tovar2009}, and Ahmed and co-workers showed
that a single oscillating bubble mixes in milliseconds~\cite{ahmed2009}. The same element has since
been used for microflow control~\cite{hashmi2013}, bidirectional pumping~\cite{gao2020pump},
micromixing on flexible substrates~\cite{lin2019mixer}, and stick-and-play pumping with single-cell
trapping~\cite{lin2019pump}. The underlying flow is the cavitation microstreaming Elder
characterized~\cite{elder1959}, whose pattern is set by the mode of bubble motion excited, pulsation, translation or shape
oscillation~\cite{doinikov2010}, by a second-order
self-interaction that multiplies the vortices around the bubble~\cite{inserra2020}, and by the presence of
neighboring bubbles~\cite{tho2007}, and which is strong enough to lyse a
vesicle~\cite{marmottant2003}, to sort particles by size~\cite{wang2011}, and to trap, pattern and
remove cells so that their interactions can be staged~\cite{faulkner2025}. The same second-order
mechanism drives the streaming around oscillating sharp edges, which are now standard microfluidic
actuators~\cite{ovchinnikov2014,nama2014}. In every case the working element is a boundary driven in oscillation,
and the quantity that matters is the steady flow this produces.

An acoustically driven gas-liquid interface supports discrete
capillary modes, and harmonic, subharmonic and superharmonic responses have been observed directly on
a meniscus in a planar micro-geometry and explained through parametric
resonance~\cite{xu2007superharmonic}. We therefore prescribe the wall motion as a single
azimuthal mode $\cos m\theta$ and ask what steady flow it drives inside the cavity.

Steady streaming has been understood since Rayleigh explained the circulation in Kundt's
tubes~\cite{rayleigh1884}, and the modern account separates a bulk mechanism, in which attenuation of
the wave supplies a non-conservative body force~\cite{eckart1948,westervelt1953,lighthill1978}, from a
boundary mechanism driven by dissipation within a thin Stokes layer at a no-slip
wall~\cite{nyborg1958}, of thickness $\delta = \sqrt{2\nu/\omega}$. Boluriaan and Morris survey the intervening
century~\cite{boluriaan2003}. Both mechanisms appear at second order in the perturbation framework
set out for microfluidics by Bruus~\cite{bruus2012} and Sadhal~\cite{sadhal2012}, and that framework
is the one used below.

Interior streaming in a closed cavity has been treated repeatedly. Nyborg derived the
near-boundary limiting streaming for a cylindrical membrane vibrating in an azimuthal
mode, treating both its interior and exterior in the thin-layer limit~\cite{nyborg1958}.
Repetto \emph{et al.}
solve the second-order flow inside a periodically rotating sphere and measure
it~\cite{repetto2008}; Sznitman and R\"osgen obtain the interior creeping-motion field of an elastic
spherical cavity, where the biharmonic equation is homogeneous and a matched boundary condition
carries the flow, which their own analysis restricts to $\delta \ll a$~\cite{sznitman2008}; Yarin computes the bulk streaming driven by shape oscillations
of a drop together with its effect on mass transfer~\cite{yarin2001}; Das \emph{et al.}
follow streaming and shape evolution together for a viscous droplet in an ultrasonic
field~\cite{das2026}; Zapryanov and Stoyanova treat
the interior and exterior streaming of a drop in translatory oscillation at small frequency
parameter~\cite{zapryanov1978}; and Kozlov and co-workers study deformable containers experimentally
across the range of $\delta/a$, both cylindrical~\cite{kozlov2017} and spheroidal with an elastic
wall~\cite{kozlov2018}. When the layer is comparable with the cavity size, they find a single
vortex pair filling the cavity. The same group reports a deformed elastic sphere under rotational
oscillations~\cite{kozlov2018pof} and a flexible container wall driven locally, where the steady rolls
sit at the nodes of the boundary oscillation~\cite{kozlov2019}. That program is an experimental one
throughout, and the analytical results are mostly exterior: Egorov and Nuriev expand the streaming
generated by an oscillating cylinder at finite frequency~\cite{egorov2021}. The peristaltic analyses
are interior ones, in a closed rectangular cavity with a travelling wave on the wall: Selverov and
Stone compute the steady streaming in such a container at high frequency and compare its Eulerian
and Lagrangian means~\cite{selverov2001}, and Yi \emph{et al.} carry a cavity with two such membrane
walls to second order in the wave amplitude and check a moving-grid finite-element code against that
expansion~\cite{yi2002}. For a fluid particle,
Doinikov \emph{et al.} give an analytical theory for streaming inside and outside under axisymmetric
shape modes of any order, with no restriction on the ratio of radius to viscous penetration
depth~\cite{doinikov2026}. Riley's survey organizes the subject around the two mechanisms
above~\cite{riley2001}. What that body of work leaves open is the case treated here: a
non-axisymmetric mode with the viscous layer filling the cavity, in closed form. One reason is
structural. Once the
Stokes layer reaches the cavity scale, the boundary-layer and core decomposition that carries the
classical closed forms loses its two regions, and an exact interior treatment has to solve one
viscous problem across the whole cavity.

An interfacial counterpart is provided by Murtsovkin and Muller. Their drop
carries an insoluble adsorption layer, which suppresses tangential motion of the surface and so
imposes zero tangential velocity exactly as a solid wall does. They compute the mean flow inside and
outside under a $P_2$ shape oscillation~\cite{murtsovkin1992}, expanding in the oscillatory
Reynolds number $U a/\nu = \varepsilon\,\mathrm{Wo}^2$, a different quantity from the
mean-flow Reynolds number of Sec.~\ref{sec:problem}, and evaluating the result for
$\beta a \gg 1$, a Stokes layer thin against the drop. The limit taken here is the opposite one,
small Womersley number $\mathrm{Wo} = a\sqrt{\omega/\nu}$, and
this limit simplifies the first-order field, as Sec.~\ref{sec:wo0} shows: at $\mathrm{Wo}^2 \to 0$ the first-order field is real
and polynomial in $r$, so the Reynolds forcing reduces to the single monomial
$r^{2m}\sin 2m\theta$ and the biharmonic inversion runs term by term with a denominator that never
vanishes, so the forcing and the resulting streamfunction are polynomials. Thin-layer descriptions instead separate a viscous boundary layer from an outer
core and determine the core flow through effective boundary conditions.

What the present problem adds to that list is a \emph{two-dimensional} disk, a
\emph{non-axisymmetric} azimuthal mode, the \emph{thick-layer} limit, and closed
forms elementary in $m$ for both the no-slip and the free-surface boundary. The whole-cavity
route is what makes them available: one biharmonic problem on the disk. Relaxing the thin
layer is known to matter.
Hamilton \emph{et al.} derived the mean mass-transport velocity for standing-wave streaming in a
two-dimensional channel of arbitrary width, needing
only that the layer be thin against the \emph{wavelength}, and found the inner boundary-layer vortices
growing relative to the Rayleigh vortices as the channel narrows, until below a width of about ten
boundary-layer thicknesses the Rayleigh vortices disappear and only the inner vortices
remain~\cite{hamilton2003}. Rednikov and
Sadhal generalized the inner streaming at a motionless boundary and the effective slip velocity it
presents to the bulk~\cite{rednikov2011}. Neither boundary deforms, and the thinness Hamilton
\emph{et al.} require is against the wavelength rather than against the channel; their mean is the
mass-transport velocity, which in a compressible channel differs from the Lagrangian mean near a
wall. Of the closest interior case, Davidson and Riley's
oscillating spherical cavity, Riley notes that they \emph{ignore the motion in the cavity}~\cite{riley2001}. The
opposite extreme, $\delta \gg a$, where the Stokes layer exceeds the cavity and the distinction
between bulk and boundary layer ceases to apply, lies outside the reach of thin-boundary-layer
descriptions built on a separate inviscid outer region.

That extreme is not exotic. Write $\mathrm{Wo} = a\sqrt{\omega/\nu}$ for the Womersley number, the
cavity radius measured in Stokes layers, so that $\delta/a = \sqrt2/\mathrm{Wo}$. The governing
parameter is its square,
$\mathrm{Wo}^2 = \omega a^2/\nu$, and viscous working
fluids are standard in this field.
Table~\ref{tab:regime} lists representative values. Its last column is the cavity radius at which
$\mathrm{Wo}^2 = 1$ at $100$ kHz, a reference scale rather than the asymptotic condition
$\mathrm{Wo}^2 \ll 1$, and the $\mathrm{Wo}^2$ columns show how fast the limit is left behind as the
cavity grows. Air is listed for comparison rather than as a candidate working fluid;
Sec.~\ref{sec:scope} gives what separates a gas bubble from this model.

\begin{table*}[t]
\centering
\caption{Representative Womersley numbers for selected fluids and cavity sizes. Kinematic
viscosities are at $20^\circ$C, except blood plasma at $37^\circ$C; glycerol concentrations are
mass fractions.}
\label{tab:regime}
\begin{tabular}{lrrrrr}
\toprule
& & \multicolumn{3}{c}{$\mathrm{Wo}^2 = \omega a^2/\nu$ at $f = 100$ kHz} & reaches \\
\cmidrule(lr){3-5}
fluid & $\nu$ [m$^2$/s] & $a = 3\,\mu$m & $a = 10\,\mu$m & $a = 30\,\mu$m
& $\mathrm{Wo}^2 = 1$ at \\
\midrule
water            & $1.0\times10^{-6}$ & $5.7$   & $63$    & $565$ & $1.3\,\mu$m \\
blood plasma     & $1.3\times10^{-6}$ & $4.3$   & $48$    & $435$ & $1.4\,\mu$m \\
50\% glycerol    & $6.0\times10^{-6}$ & $0.94$  & $10.5$  & $94$ & $3.1\,\mu$m \\
air              & $1.51\times10^{-5}$ & $0.37$  & $4.2$   & $37$ & $4.9\,\mu$m \\
85\% glycerol    & $1.0\times10^{-4}$ & $0.057$ & $0.63$  & $5.7$ & $12.6\,\mu$m \\
glycerol         & $1.18\times10^{-3}$& $0.0048$& $0.053$ & $0.48$ & $43.3\,\mu$m \\
\bottomrule
\end{tabular}
\end{table*}

\section{Problem}
\label{sec:problem}

Let the unit disk contain an incompressible Newtonian fluid whose boundary oscillates at angular
frequency $\omega$ with small amplitude. To leading order the response is an oscillatory Stokes flow
$\mathbf{u}_1 = \mathrm{Re}[\hat{\mathbf{u}}_1 e^{-i\omega t}]$ satisfying
$-i\,\mathrm{Wo}^2 \hat{\mathbf{u}}_1 = -\nabla \hat p_1 + \nabla^2 \hat{\mathbf{u}}_1$ with
$\nabla\cdot\hat{\mathbf{u}}_1 = 0$. In the limit $\mathrm{Wo}^2 = 0$,
\begin{equation}
\nabla \hat p_1 = \nabla^2 \hat{\mathbf{u}}_1, \qquad
\hat u_{1r}(1,\theta) = a_m \cos m\theta, \qquad \hat u_{1\theta}(1,\theta) = 0 .
\label{eq:first}
\end{equation}
The boundary data carries zero net flux for every $m \ge 1$, so no flux correction is needed at first
order. It is needed at second order. A material wall at $R(\theta,\tau) = 1 + \epsilon\sin\tau\cos m\theta$ encloses an area
$\pi[1 + \tfrac12\epsilon^2\sin^2\tau]$, which is not constant, so an incompressible closed cavity
requires a second-order correction to the wall motion; adding an axisymmetric $\epsilon^2 R_2(\tau)$
with $R_2 = -\tfrac14\sin^2\tau$ restores the area to this order. That correction carries a $2\omega$
radial velocity and no mean, so it does not enter the steady problem of Eq.~\eqref{eq:second} and leaves
Eqs.~\eqref{eq:main} and~\eqref{eq:free} unchanged, though a moving-mesh computation must include
it, so that the instantaneous cavity area is preserved through second order. With $\bm{\tau}_R = \tfrac12 \mathrm{Re}(\hat{\mathbf{u}}_1
\otimes \hat{\mathbf{u}}_1^{*})$ the steady streaming obeys
\begin{equation}
-\nabla^2 \bar{\mathbf{u}} + \nabla p_2 = -\nabla \cdot \bm{\tau}_R, \qquad
\nabla \cdot \bar{\mathbf{u}} = 0, \qquad \bar{\mathbf{u}}|_{r=1} = \mathbf{0}.
\label{eq:second}
\end{equation}
At $\mathrm{Wo}^2 = 0$ the first-order field is exactly real, so $\bm{\tau}_R$ is a rank-one dyad and
the forcing stays in closed form.

The mean fields below need separating at the outset. $\psi_2(\mathrm{Wo})$ solves Eq.~\eqref{eq:second} with
$\bar{\mathbf{u}}|_{r=1}=\mathbf{0}$, an \emph{auxiliary} problem in which no slip is imposed on the
Eulerian mean; this is what a second-order solver conventionally poses and what
Sec.~\ref{sec:verify} checks. Nama \emph{et al.}~\cite{nama2017} label that choice
$\mathrm{E}_{\mathrm{zbc}}$, homogeneous Dirichlet data on the second-order Eulerian velocity, and
take Muller \emph{et al.}~\cite{muller2012} as their example of it; they set it against an Eulerian
formulation with moving-boundary conditions, which they label $\mathrm{E}_{\mathrm{mbc}}$, and
against their own arbitrary Lagrangian--Eulerian formulation, which carries the
Lagrangian mean as the unknown of the second-order problem and so takes a homogeneous wall value
without any Stokes-drift post-processing. Their separation of scales is a perturbation expansion,
as here; the moving-domain computation of Sec.~\ref{sec:verify} is instead a transient nonlinear
solve, and the two are complementary rather than alternatives. Its $\mathrm{Wo}^2\to0$ limit is Eq.~\eqref{eq:main}, written
$\psi_2$, and $\psi_2^{\,\mathrm{f}}$ is that same limit with a free surface, Eq.~\eqref{eq:free}.
$\psi_E$ is the Eulerian mean of the physical moving-wall problem, whose mean velocity
$\bar{\mathbf{u}}_E$ carries the generally nonzero wall value
$\bar{\mathbf{u}}_E(1) = -\langle(\bm{\xi}_1\!\cdot\!\nabla)\mathbf{u}_1\rangle
= -\tfrac12\mathrm{Re}\,\langle(\hat{\bm{\xi}}_1^{*}\!\cdot\!\nabla)\hat{\mathbf{u}}_1\rangle$,
with $\bm{\xi}_1$ the first-order particle displacement and $\hat{\bm{\xi}}_1$ its complex
amplitude. $\psi_S$ is the Stokes drift, and $\psi_L = \psi_E + \psi_S$ is the
Lagrangian mean, the field a tracer follows. Section~%
\ref{sec:lagrangian} gives $\psi_L$ and shows it differs from $\psi_2$ by a fixed factor even as
$\mathrm{Wo}^2\to0$.

Two conditions are assumed throughout. Dropping
$(\bar{\mathbf{u}}\!\cdot\!\nabla)\bar{\mathbf{u}}$ from Eq.~\eqref{eq:second} requires the mean flow's own
Reynolds number to be small. That number depends on the scale of the mean field, not only on the
oscillation: with $\epsilon = U/(\omega a)$ the displacement amplitude scaled by the radius and $F$ the
dimensionless size of $\bar{\mathbf{u}}$ in the scale $U^2a/\nu$ set by Eq.~\eqref{eq:second},
$\mathrm{Re}_{\mathrm{mean}} = \epsilon^2\mathrm{Wo}^4 F$. In the viscous-dominated limit $F = O(1)$
and the condition is $\epsilon^2\mathrm{Wo}^4 \ll 1$, satisfied automatically as
$\mathrm{Wo}\to 0$ at fixed small $\epsilon$. At the top of the
$\mathrm{Wo}$ range Sec.~\ref{sec:finiteWo} reaches, the auxiliary deforming-wall field
$\psi_2(\mathrm{Wo})$ has $F = O(\mathrm{Wo}^{-1})$ and $\psi_L$ is smaller by a further power of
$\mathrm{Wo}$; taking the larger, the requirement is $\epsilon^2\mathrm{Wo}^3 \ll 1$, and at
$\mathrm{Wo}^2 = 10^5$ that already demands $\epsilon \lesssim 10^{-4}$.
A wall condition applied on the reference circle rather than the displaced one further requires
$\epsilon\,\mathrm{Wo}/\sqrt2 \ll 1$, the displacement measured against the Stokes layer rather than
the radius, and $\epsilon m \ll 1$, measured against the radial scale $a/m$ of a mode-$m$ field. The
last is what the boundary slope $O(\epsilon m)$ costs, and it is the condition that makes the
large-$m$ laws below a joint limit $1 \ll m \ll \epsilon^{-1}$ rather than a statement at fixed
displacement. Of the two, the mean-flow condition binds: at $\mathrm{Wo}^2 = 10^5$ and
$\epsilon = 10^{-4}$ the Stokes-layer ratio is $0.02$, while
$\epsilon^2\mathrm{Wo}^3$ is already $0.3$. The high-$\mathrm{Wo}$ end of Sec.~\ref{sec:finiteWo} therefore
reads as an asymptotic check on the construction, valid at very small amplitude.
Incompressibility requires the cavity to be small against the acoustic wavelength
$\lambda_{\mathrm{ac}} = c/f$, and every entry of Table~\ref{tab:regime} is acoustically compact: the
least favourable case, a $30\,\mu$m cavity in air at $100$ kHz, still has
$\lambda_{\mathrm{ac}}/a \approx 114$; it also removes wave attenuation, and with it the Eckart
mechanism, so \emph{bulk} is used throughout in the spatial sense of the cavity interior rather than
for the attenuation-driven mechanism named in Sec.~\ref{sec:intro}.

\section{Solution}
\label{sec:wo0}

With $\psi_1$ the first-order streamfunction, Stokes flow makes it biharmonic; the solutions regular
at the origin are $r^m$ and $r^{m+2}$, and the two conditions in Eq.~\eqref{eq:first} fix both:
\begin{equation}
\psi_1 = a_m\left[\frac{(m+2)\,r^{m}}{2m} - \frac{r^{m+2}}{2}\right]\sin m\theta .
\end{equation}
Taking the curl of Eq.~\eqref{eq:second} removes the pressure and every gradient part of the forcing. With
$\bar{\mathbf{u}} = \nabla \times (\psi_2 \hat{\mathbf{z}})$ we need
$\nabla^4 \psi_2 = \nabla\times(-\nabla\cdot\bm{\tau}_R)$ subject to
$\psi_2 = \partial_r\psi_2 = 0$ at $r=1$. Since $\hat u_{1r} \propto \cos m\theta$ and
$\hat u_{1\theta} \propto \sin m\theta$, the quadratic products carry only harmonics $0$ and $2m$, and
the axisymmetric part is purely radial, hence a gradient absorbed into $p_2$ that drives no flow. For
$a_m = 1$,
\begin{equation}
\nabla \times (-\nabla \cdot \bm{\tau}_R) = -\frac{m(m+1)}{2}\, r^{2m} \sin 2m\theta ,
\end{equation}
which the $r^{m+2}$ part of $\psi_1$ supplies on its own. The $r^m$ part is harmonic, hence
irrotational, and for an irrotational incompressible field $(\mathbf{u}\cdot\nabla)\mathbf{u}$ is a
pure gradient, so it contributes nothing to the curl. The cross terms go with it, because the
vorticity $-\nabla^2\psi_1$ is proportional to the $r^m$ harmonic itself and is therefore constant
along the streamlines of the $r^m$ part: writing $\psi_1 = (A r^m + B r^{m+2})\sin
m\theta$, the forcing is $-2m(m+1)B^2 r^{2m}\sin 2m\theta$, a function of $B$ alone. The streaming
amplitude therefore traces to the vortical component of the oscillatory field.
Since $\nabla^4(r^k \sin n\theta) = (k^2-n^2)((k-2)^2-n^2)\,r^{k-4}\sin n\theta$, the inversion is
term by term with denominator $64(m+1)(2m+1)$, which never vanishes. Adding $r^{2m}$ and $r^{2m+2}$ and
imposing both no-slip conditions,
\begin{equation}
\psi_2 = -\frac{m}{128(2m+1)}\; r^{2m}\left(r^2-1\right)^2 \sin 2m\theta \,.
\label{eq:main}
\end{equation}
The factor $(r^2-1)^2$ carries both no-slip conditions.

\begin{figure*}[t]
\centering
\includegraphics[width=\textwidth]{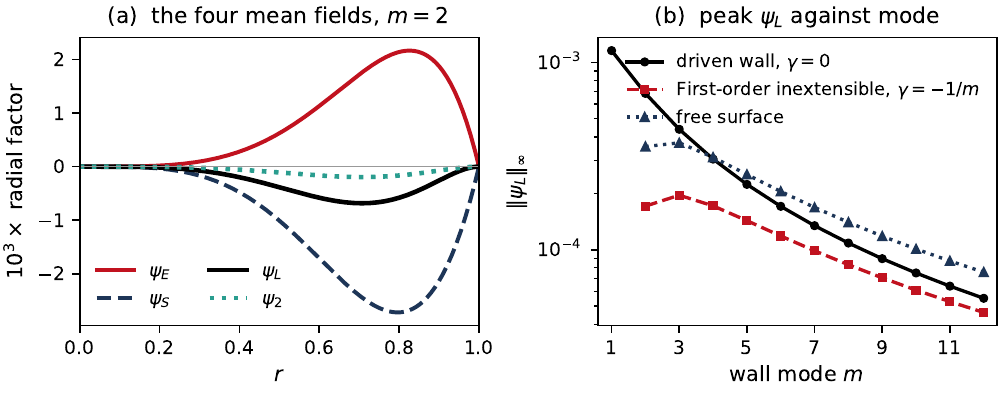}
\caption{(a) The four mean fields at $m=2$, as radial profiles of the streamfunction at the angle of
the cell center. The physical Eulerian mean $\psi_E$, Eq.~\eqref{eq:eulerian}, and the Stokes drift
$\psi_S$, Eq.~\eqref{eq:stokesdrift}, are each an order of magnitude larger than the
reference-boundary field $\psi_2$, Eq.~\eqref{eq:main}, and nearly cancel; what survives is the
Lagrangian mean $\psi_L$, Eq.~\eqref{eq:lagrangian}, the field a tracer follows. $\psi_E$ circulates
opposite to $\psi_2$. (b) Peak $\|\psi_L\|_\infty$ against mode number for three wall kinematics: an
externally driven extensible wall falls monotonically from $m=1$ at fixed radial
wall-velocity amplitude, while the first-order
surface-inextensible family, $\gamma=-1/m$, and a shear-free interface both vanish at $m=1$ and
peak at $m=3$. The ranking is a property of
the wall, not of the cavity.}
\label{fig:means}
\end{figure*}

Equation~\eqref{eq:main} solves the reference-boundary problem, with the Eulerian mean set to zero
on $r=1$. The physical moving wall leaves the slip of Eq.~\eqref{eq:sliplimit} instead, and carrying
that through the same inversion gives the Eulerian mean of the cavity in closed form,
\begin{equation}
\psi_E = -\frac{r^{2m}(r^2-1)\left[(m^2+2m)r^2 - m^2 + 14m + 8\right]}
              {128\,(m+2)(2m+1)}\,\sin 2m\theta ,
\label{eq:eulerian}
\end{equation}
which satisfies $\psi_E(1,\theta) = 0$ and the slip of Eq.~\eqref{eq:sliplimit} at the wall. It is
the field an experiment measuring the Eulerian mean would see, and it stands in a different relation
to Eq.~\eqref{eq:main} than a correction does: at $m=2$ and $\gamma = 0$ its peak streamfunction
amplitude is $11.1$ times that of Eq.~\eqref{eq:main}, with the opposite sign, so the circulation
runs the other way. That number compares two streamfunction peaks at one mode and one wall
kinematics; it is not a ratio of speeds and not $\psi_E/\psi_L$. Figure~\ref{fig:means}(a) shows the four mean fields together:
$\psi_E$ and $\psi_S$ are both large and nearly opposite, and $\psi_L$ is what is left of them.
Equation~\eqref{eq:main} remains what a second-order solver returns
if it is given the conventional homogeneous condition on the reference circle, which is why
Sec.~\ref{sec:verify} tests it against one, and Sec.~\ref{sec:lagrangian} assembles
$\psi_L = \psi_E + \psi_S$ from it. A solver given the material wall instead returns $\psi_E$.

\subsection{Properties of the small-Womersley-number solution}

\emph{Spatial frequency doubling.} A wall deforming at azimuthal mode $m$ drives streaming at
$2m$: an elliptical $m=2$ mode produces \emph{eight} counter-rotating cells, not four
[Fig.~\ref{fig:fields}(a)]. The quadratic self-interaction of a mode-$m$ first-order field
produces an azimuthal harmonic of order $2m$, and hence $4m$ alternating cells. Superharmonic
response is also seen in the time domain on driven menisci~\cite{xu2007superharmonic}. The Rayleigh slip problem doubles in the same way: its
limiting slip goes as $\sin 2m\theta$ too, so the classical four-vortex pattern is its $m=1$ case.

\emph{Where the cells sit.} Differentiating $r^{2m}(r^2-1)^2$ gives the streamfunction
extremum, that is the cell center, at
\begin{equation}
r_{\max} = \sqrt{\frac{m}{m+2}},
\end{equation}
which is $0.577$ at $m=1$, $0.707$ at $m=2$, and $0.913$ at $m=10$. Higher drive modes push the
streaming against the wall and leave the interior progressively unstirred.

\emph{The prefactor saturates but the field does not.} The coefficient $-m/(128(2m+1))$ tends to
$-1/256$, but the radial factor shrinks at the same time. Evaluating Eq.~\eqref{eq:main} at its own extremum
$r_{\max} = \sqrt{m/(m+2)}$,
\begin{equation}
\|\psi_2\|_\infty = \frac{m^{m+1}}{(m+2)^{m+2}\,32(2m+1)}
\;\sim\; \frac{e^{-2}}{64\,m^{2}} \qquad (m \to \infty),
\label{eq:peak}
\end{equation}
which is $1/2592$, $1/5120$, $81/700000$, $1/13122$ at $m = 1,2,3,4$. The decrease holds at every
$m$ rather than over a sampled range: $\mathrm{d}\log\|\psi_2\|_\infty/\mathrm{d}m =
\log[m/(m+2)] + 1/[m(2m+1)]$, which is $-0.765$ at $m=1$ and at most $(5-4m)/2m^2$ thereafter.
That the decrease is monotone is exact; the $m^{-2}$ rate is the limit it approaches, and it
approaches slowly. Writing $\alpha_m = \log(\|\psi_2\|_\infty^{(m)}/\|\psi_2\|_\infty^{(m-1)})
/\log[m/(m-1)]$ for the exponent between successive modes, $\alpha_m$ is $-1.291$ at $m=3$,
$-1.621$ at $m=6$ and $-1.939$ at $m=40$, with $m^2\|\psi_2\|_\infty \to e^{-2}/64$. Over the mode
numbers a device would use, the exact peak of Eq.~\eqref{eq:peak} is the thing to quote.
The Lagrangian factor of Sec.~\ref{sec:lagrangian}, which tends to $5$, does not change this.
The peak speed follows from the same two factors. Writing
$A_m = m(5m+4)/[128(m+2)(2m+1)]$, the radial component of the Lagrangian mean is
$u_{Lr} = -2mA_m r^{2m-1}(1-r^2)^2\cos 2m\theta$ at $a_m = 1$, whose magnitude peaks at
$r^2 = (2m-1)/(2m+3)$ and is asymptotic to $5e^{-2}/(32m)$. The azimuthal component is bounded at
the same order, so the peak speed falls as $m^{-1}$. It approaches even more slowly than the
$m^{-2}$ law: at $m=2$ the exact peak speed is $0.00401$ against an asymptote of $0.0106$. For
small $m$ the exact expression is again the thing to use. Any
statement of an optimal mode requires a target and a constraint to be named first, since fixing the
wall velocity, the displacement, the power or the wall stress gives a different optimization in each
case. On this measure there is no interior optimum: raising $m$ makes the streaming weaker
\emph{and} confines it nearer the wall.

\emph{The same field in physical units.} The scale $U^2a/\nu$ hides how slow this flow is. Take the glycerol row of Table~\ref{tab:regime} at $a = 10\,\mu$m and $f = 100$ kHz, so
$\mathrm{Wo}^2 = 0.053$, and drive the wall at one percent of the radius. Then $U = \epsilon\omega a
= 6.3$ cm/s, the streaming scale is $33.5\,\mu$m/s, and the peak speed of the Lagrangian field at
$m=2$, which is $0.00401$ in that scale, is $0.134\,\mu$m/s. A tracer at that speed covers one cavity
radius in $75$ s. That is a transport time. Mixing is a further question, because the cells are
closed and transfer between neighboring vortices is diffusive; a cell count is a statement about
the flow pattern rather than about how fast anything homogenizes. For a small molecule the two
compete evenly: with a hydrodynamic radius of $0.30$ nm the diffusivity in glycerol is
$4.8\times10^{-13}\,$m$^2$/s and the P\'eclet number $u_L a/D$ is $2.8$: advection and diffusion carry a small
molecule across the cavity at comparable rates. Whether a given initial distribution homogenizes
faster is a different question, needing a concentration problem and a chosen measure of mixedness,
and this number does not answer it. What it does transport is particles. A $1\,\mu$m
polystyrene bead follows the flow some $1700$ times faster than it drifts under its own buoyancy,
and in glycerol it drifts \emph{upward}, being the lighter phase.

\emph{Multi-mode drives, and why the inversion always closes.} For a superposition the products
carry $2m$, $2m'$, $m+m'$ and $|m-m'|$. The second-order problem is linear in the forcing, so
harmonics do not interact and each is inverted with its own pair of no-slip constants. The apparent
hazard is a vanishing denominator, which would require a $\log r$ particular solution. It cannot
occur, for any drive and any $m$.

Write the forcing as a sum of terms $r^p \sin n\theta$. Inversion takes $k = p+4$, so
$D = ((p+4)^2-n^2)((p+2)^2-n^2)$ vanishes only when $n = p+2$ or $n = p+4$, that is only when $n > p$.
The combination is not hypothetical. Mode $m$ builds $\hat{\mathbf{u}}_1$ from $r^{m-1}$ and
$r^{m+1}$, so a product of modes $m$ and $m'$ carries radial powers $m+m'-2$, $m+m'$ and $m+m'+2$;
the curl of the divergence lowers the power by two, so $p$ ranges over $m+m'-4$, $m+m'-2$, $m+m'$.
Setting $n = m+m'$ places $p = n-4$ and $p = n-2$ exactly on the two resonant values, and with
$\min(m,m') = 1$ the cross harmonic $|m-m'| = m+m'-2$ lands on $p = n-2$ as well.

Those coefficients vanish identically. Since
$r^m \sin m\theta = \mathrm{Im}(z^m)$ and $r^{m+2}\sin m\theta = r^2\,\mathrm{Im}(z^m)$, the
first-order streamfunction is a \emph{polynomial} in $x$ and $y$; hence so are
$\hat{\mathbf{u}}_1$, $\bm{\tau}_R$ and $\nabla\times(\nabla\cdot\bm{\tau}_R)$. A homogeneous
polynomial of degree $p$ contains only harmonics $n \le p$ with $n \equiv p \pmod 2$, so a term
$r^p \sin n\theta$ with $n > p$ cannot occur in a homogeneous Cartesian polynomial of degree $p$.
Degeneracy of the term-by-term inversion requires $n > p$; the polynomial structure forbids it. The
inversion is therefore nondegenerate for any number of drive modes and any $m$. This is a statement
about the monomials of the biharmonic inversion, not about acoustic, capillary or elastic
resonances of a device. Polynomial existence and uniqueness support the term-by-term inversion. Karachik and Antropova construct polynomial solutions of the inhomogeneous biharmonic
Dirichlet problem with polynomial data in a ball~\cite{karachik2013}, which is what makes the
term-by-term inversion legitimate. Li and Ponnusamy give the biharmonic Green function of the unit
disk and prove that the clamped problem $\nabla^4 w = g$ with $w = \partial_n w = 0$ on the boundary
has exactly one solution for $g$ continuous on the closed disk~\cite{liponnusamy2017}. The forcing
here is a polynomial, so Eq.~\eqref{eq:main} is that solution, and a solver of Eq.~\eqref{eq:second} has a
single field to converge to. The statement is tied to the clamped pair of conditions. A free surface
replaces the second of them by a vanishing tangential stress, a second-derivative condition in a
different fourth-order class, so uniqueness for Eq.~\eqref{eq:free} is a separate question and this
citation does not settle it, and the answer is concrete: the free-surface problem has a nullspace.
For any constant $C$, the field $C(r^2-1)$ satisfies $\nabla^4\psi = 0$, vanishes on $r=1$ and
carries no tangential stress there, and corresponds to a rigid rotation $u_\theta = -2Cr$. A solver
run against Eq.~\eqref{eq:free} must therefore fix $C$, by requiring zero total angular momentum or
by solving in the forced $n = 2m$ subspace alone; Eq.~\eqref{eq:free} is the latter choice.

\emph{A single standing mode carries no axisymmetric Reynolds shear at any $\mathrm{Wo}$, and we
take the branch with no mean rotation.} The two
agree. A single standing mode has no axisymmetric Reynolds shear stress at any Womersley
number: its angular dependence is proportional to $\cos m\theta\,\sin m\theta$, whose azimuthal
average vanishes for any complex radial amplitude. We select the zero-angular-momentum branch,
$C = 0$. The complete torque balance on the displaced wall also includes pressure and viscous
traction. The reason the Reynolds term vanishes is angular rather than dynamical, which makes it
exact at every $\mathrm{Wo}$ and specific to a standing mode. In the thin-layer exterior limit the
corresponding statement is classical: for a rigid ($m=1$) circular cylinder whose center moves on an
elliptic path of axis ratio $\lambda$, the axisymmetric part of Riley's slip velocity is carried by
$\lambda$ alone and vanishes for the transverse vibration $\lambda = 0$~\cite{riley1998}.
Longuet-Higgins's analysis of an island in an oscillating current makes the same
split~\cite{lh1970}, and Rednikov and Sadhal obtain it on a sphere~\cite{rednikov2011}. Those are thin-layer
results for an external flow and do not by themselves establish the interior statement above. It is also
specific to this domain: the orthogonality is an integral around a closed circle, and the rotation
it excludes is the one rigid mode a simply connected two-dimensional cavity admits. An
axisymmetric three-dimensional cavity admits meridional circulation that no azimuthal average
constrains. A
$\theta$-independent drive needs no such orthogonality: continuity makes $r\,u_r$ independent of
$r$, regularity at the origin then forces $u_r \equiv 0$, and the axisymmetric Reynolds shear
vanishes identically whatever the phase of $f$.

One restriction applies to the multi-mode case. The first-order field is real at
$\mathrm{Wo}^2 = 0$ only when all driven modes share a time phase. With a relative phase, for instance
$\cos 2\theta + i\cos 3\theta$, the Stokes drift is nonzero even in the inertialess limit, and the
low-$\mathrm{Wo}$ reading of Sec.~\ref{sec:lagrangian} does not carry over; the polynomial structure
and the non-degeneracy of the inversion are unaffected.

The inversion was checked symbolically on seventeen mode pairs, chosen to cover each structure the
power counting distinguishes: equal modes, where the forcing is the single harmonic $2m$;
$\min(m,m') = 1$, where the cross harmonic $|m-m'|$ lands on a resonant slot as well; and unequal
modes both adjacent and separated, out to $m = 13$. In every case the predicted resonant slots are
empty, every term present satisfies $n \le p$ with matching parity, and the smallest $|D|$ encountered
is $192$. A planted $r^{0}\sin 4\theta$, for which $D = 0$, is flagged, so these are the passes of a
detector that fires.

\subsection{What the wall kinematics decides}
\label{sec:gamma}

Equation~\eqref{eq:main} fixes the tangential wall velocity to zero. A wall can instead carry one,
and the ranking over $m$ turns out to depend on which. Prescribe both components at the same time
phase,
\begin{equation}
\hat u_{1r}(1,\theta) = a_m\cos m\theta, \qquad
\hat u_{1\theta}(1,\theta) = \gamma\,a_m \sin m\theta ,
\label{eq:gammabc}
\end{equation}
with $\gamma$ real. The two conditions give $B = -(1+\gamma)/2$ in
$\psi_1 = (A r^m + B r^{m+2})\sin m\theta$, and although $f$ itself is not proportional to
$1+\gamma$, the resonant coefficient of the Reynolds forcing is proportional to its square, so
\begin{equation}
\psi_2(\gamma) = (1+\gamma)^2\,\psi_2(0) ,
\label{eq:gammascale}
\end{equation}
and Eq.~\eqref{eq:lagrangian} scales with it. The Lagrangian factor $(5m+4)/(m+2)$ is unchanged,
because it is a ratio and $(1+\gamma)^2$ is common to both fields.

The square has a reason. The first-order vorticity of Eq.~\eqref{eq:gammabc} is
$\omega_1 = 2(m+1)a_m r^m(1+\gamma)\sin m\theta$, so $1+\gamma$ is the amplitude of the vorticity
the wall injects, and Eq.~\eqref{eq:gammascale} says the streaming is quadratic in it. At
$\gamma = -1$ the first-order field is irrotational, its nonlinear term is a gradient absorbed into
the pressure, and the curl of the Reynolds forcing vanishes at every $m$.

Two prescribed members of this family illustrate the role of material motion. At $\gamma = 0$ the
wall has no first-order tangential velocity and its material stretches azimuthally, and it is
Eq.~\eqref{eq:main}. At $\gamma = -1/m$ the linearised surface strain rate vanishes,
\begin{equation}
\partial_\theta \hat u_{1\theta} + \hat u_{1r} = 0 \qquad (r = 1),
\label{eq:strain}
\end{equation}
with amplitude factor $[(m-1)/m]^2$. We call this \emph{first-order surface-inextensible
kinematics}, and refer to it below as the $\gamma = -1/m$ family. It is a prescribed kinematics,
not a general statement about an inextensible material. We complete the motion by conserving the
enclosed area and allowing second-order extension; for an initially circular boundary and a
nontrivial mode $m \ge 2$, area and perimeter cannot both stay constant through second order.
Appendix~\ref{app:material} gives that completion explicitly and verifies that the material motion
is periodic with no net displacement over a cycle, so it adds no independent mean wall velocity.
The factor $[(m-1)/m]^2$ vanishes at $m=1$, where the first-order motion is a rigid translation,
and it rises with $m$ while Eq.~\eqref{eq:peak} falls, so at fixed radial wall-velocity amplitude
the product peaks at $m=3$ rather than at $m=1$ [Fig.~\ref{fig:means}(b)]. The mode that stirs
hardest is therefore a property of the wall and not of the cavity.

Which mode is best also depends on what is held fixed, and the comparisons here hold the
\emph{radial} wall-velocity amplitude $a_m$ constant. Holding the cycle- and circumference-averaged
root-mean-square wall speed constant instead rescales $a_m$ by $(1+\gamma^2)^{-1/2}$, which for
the $\gamma = -1/m$ family is $m/\sqrt{m^2+1}$ and so varies with the mode; computed over $m \le 12$ it moves
neither ranking. Holding the cycle-averaged power constant does move one. That power is
$\pi(m\gamma^2 + 2\gamma + m)$ in $\mu U^2$ per unit depth, so at $\gamma = 0$ it rises with $m$ and
$a_m^2$ falls as $1/m$; the externally driven wall is still largest at $m=1$, and the
$\gamma = -1/m$ optimum moves from $m=3$ to $m=2$. Holding the first-order wall shear constant does
not apply to that family at all: that stress is $2|m\gamma+1|a_m$, which vanishes identically at
$\gamma = -1/m$, because a wall with no first-order surface strain exerts no first-order tangential
traction. Under each of these the $\psi_2$ and $\psi_L$ rankings agree, so the choice of mean field
does not enter.

At $m=1$ two members of the family are the same shape and different motions, which is the
sharpest form of that statement. A disk translating rigidly has $\hat u_{1r}(1,\theta) = a_1\cos\theta$ with
$\hat u_{1\theta}(1,\theta) = -a_1\sin\theta$, so it is $\gamma = -1$ and drives nothing, as a rigid
motion of the whole cavity must. Holding the tangential velocity at zero instead, $\gamma = 0$,
leaves the boundary tracing the same circle to first order while its material stretches
azimuthally, and it drives the Lagrangian mean of Eq.~\eqref{eq:lagrangian} in full. The geometry
of the
moving boundary does
not determine the mean flow; the motion of the wall material does. Taylor's analysis of a swimming
sheet rests on the same distinction: the shape of the sheet fixes only the velocity normal to it,
and a further assumption about the sheet, which he takes to be inextensibility, is needed to fix
the tangential one~\cite{taylor1951}.

These are two prescribed kinematics, not a bound on elastic walls. A shell with a constitutive
response can put its tangential motion out of phase with its radial motion, which
Eq.~\eqref{eq:gammabc} does not allow, and settling where a given material sits needs a shell model
this paper does not carry.

\subsection{The same disk with a free surface}

Acoustically driven sessile droplets are a standard circular geometry in this
field, with the free surface itself used to confine the sound~\cite{yu2011}, and there the tangential
\emph{stress} vanishes rather than the tangential velocity. Two-dimensional streaming at a compliant boundary is well studied on the \emph{outside}. Rallabandi \emph{et al.} give an asymptotic theory for the flow around a sessile semicylindrical bubble and
show that mixed-mode streaming explains the robustness of the fountain pattern~\cite{rallabandi2013}.
Bhosale \emph{et al.} identify a distinct streaming process driven by boundary elasticity around an
immersed soft cylinder, ``available even in Stokes flows''~\cite{bhosale2022}, and Cui \emph{et al.} carry
it to three dimensions~\cite{cui2024}. That mechanism is distinct from the one here: it needs an
elastic constitutive response, which this prescribed-boundary problem does not model. The mean flow
here is driven by the convective Reynolds stress, and its dimensional scale
$U^2a/\nu = \varepsilon^2\omega a\,\mathrm{Wo}^2$ vanishes with $\mathrm{Wo}^2$ at fixed
displacement amplitude and frequency. What survives the limit is the coefficient after that
scaling, so this is the leading weak-inertia mean rather than rectification without inertia. That
holds for a standing wall mode. This paper treats a single standing wall mode. Each of these solves the region outside an
immersed feature, where the biharmonic admits $\ln r$ and $r^{-2}$. Of the four homogeneous radial
solutions of $\nabla^4[f(r)\sin n\theta] = 0$, namely $r^{n}$, $r^{n+2}$, $r^{-n}$ and $r^{2-n}$ for
$n \ge 2$, regularity at the origin excludes the last two inside the circle, so the homogeneous
space has dimension two where the exterior has four. The two survivors and the particular term at
$r^{2m+4}$ make Eq.~\eqref{eq:free} a three-term polynomial. The field $\psi_1^{\,\mathrm{f}}$
below has two terms for $m \ge 2$, collapsing to $r$ at $m=1$: at first order there is no forcing,
so the field is purely homogeneous. Replacing $\hat u_{1\theta}(1,\theta) = 0$
in Eq.~\eqref{eq:first} by $\hat\sigma_{r\theta}(1,\theta) = 0$, and changing nothing else, gives
\begin{equation}
\psi_1^{\,\mathrm{f}} = \frac{(m+1)r^{m} - (m-1)r^{m+2}}{2m}\,\sin m\theta ,
\end{equation}
a forcing $-[(m-1)^2(m+1)/2m]\,r^{2m}\sin 2m\theta$ that is again a pure monomial, and
\begin{equation}
\psi_2^{\,\mathrm{f}} = -\frac{(m-1)^2}{128\,m^2(2m+1)}\;
r^{2m}\left(r^2-1\right)\left(m r^2 - m - 1\right)\sin 2m\theta \,.
\label{eq:free}
\end{equation}
The mean conditions used here, $\psi_2^{\,\mathrm{f}}(1,\theta) = 0$ and zero mean tangential
stress, define a \emph{reference-boundary} problem on the circle $r=1$. They are not the physical
conditions on the displaced interface. Imposing the interface conditions at $r = 1+\xi$ and
projecting onto the rotated normal and tangent generates two second-order terms, $\langle
\xi\,\partial_r \hat\sigma_{r\theta}\rangle$ and $\langle \partial_\theta\xi\,(\hat\sigma_{rr} -
\hat\sigma_{\theta\theta})\rangle$. Evaluated on a strictly real first-order field these vanish,
because the displacement is then in quadrature with every first-order stress; but the mean is
measured in $U^2a/\nu$, and the $O(\mathrm{Wo}^2)$ phase correction to the first-order field enters
that scale divided by $\mathrm{Wo}^2$. The terms therefore survive the limit, exactly as the
second-order wall slip of Sec.~\ref{sec:problem} does.

Writing $f = f_0 + i\,\mathrm{Wo}^2 f_1 + \cdots$ for the free-surface first-order radial factor, the
correction is
\begin{equation}
f_1^{\,\mathrm{f}} = \frac{r^{m}(m-1)(r^2-1)(mr^2-m-2)}{16m^2(m+2)} .
\label{eq:f1free}
\end{equation}
Carrying it through the displaced-interface kinematic and tangential-stress conditions leaves the
physical Eulerian mean $\psi_E^{\,\mathrm{f}} = h_E(r)\sin 2m\theta$ with inhomogeneous data,
\begin{equation}
h_E(1) = \frac{m-1}{16m^2(m+2)}, \qquad
T_{2m}[h_E](1) = -\frac{(m-3)(m-1)}{8m(m+2)},
\label{eq:physfree}
\end{equation}
where $T_n[h] = -h'' + h'/r - n^2h/r^2$ is the radial factor of the tangential stress. At $m=2$ these
are $1/256$ and $1/64$. Equation~\eqref{eq:free} is therefore the reference-boundary solution, and
the physical mean follows from it in Sec.~\ref{sec:lagrangian}.

The radial factor is no longer a perfect square, because the second
condition is a stress and not a velocity; the harmonic is still $2m$, and because the extra root sits at $r = \sqrt{1+1/m} > 1$, the cell count is still $4m$. The cell center moves outward, to
$0.761$ against $0.707$ at $m = 2$. And Eq.~\eqref{eq:free} vanishes identically at $m = 1$, where a free
surface driven at $\cos\theta$ translates the disk rigidly, leaving the Reynolds stress nothing to
rectify. The free-surface reference-boundary peak $\|\psi_2^{\,\mathrm{f}}\|_\infty$ is
$0$, $1.02\times10^{-4}$, $9.80\times10^{-5}$ and $7.79\times10^{-5}$ at $m = 1,2,3,4$ and decays
as $m^{-2}$ beyond, so on that measure $m = 2$ is the largest. The physical field ranks the modes
differently. Because the Lagrangian factor $(5m+4)/(m+2)$ of Sec.~\ref{sec:lagrangian} rises with
$m$, the peak of $\psi_L^{\,\mathrm{f}}$ is $3.554\times10^{-4}$ at $m = 2$ and
$3.723\times10^{-4}$ at $m = 3$: at fixed wall-velocity amplitude the shear-free Lagrangian
streamfunction peaks $4.75\%$ higher at $m = 3$, the largest value at any $m$, since the peak
decreases monotonically for $m \ge 3$. Which mode a
device should use depends on the target and the
constraint, since peak streamfunction, peak speed, power and homogenization time are optimized
differently. Comparing reference-boundary peak magnitudes, the shear-free boundary is the weaker
driver at low $m$ and the stronger at high $m$, crossing over near $m = 4$:
the ratio is $0.52$ at $m=2$, $1.02$ at $m=4$ and $1.34$ at $m=10$.

\begin{figure*}[t]
\centering
\includegraphics[width=\textwidth]{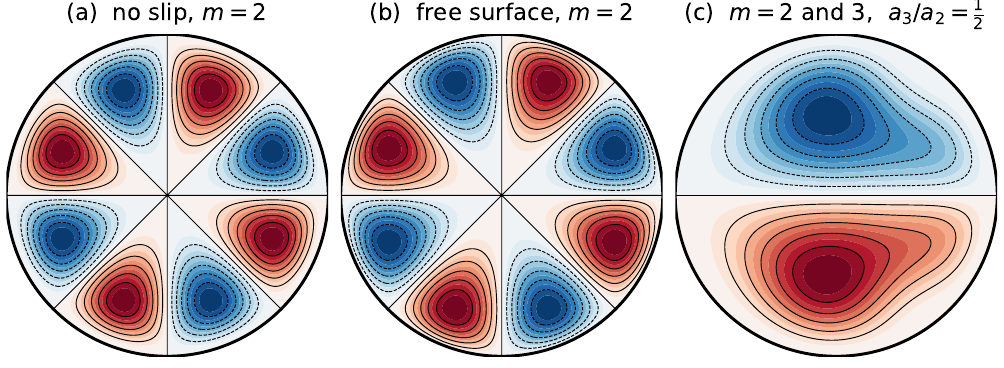}
\caption{(a) The exact no-slip streaming for $m=2$, Eq.~\eqref{eq:main}: the $\sin 4\theta$ structure
gives eight counter-rotating cells. (b) The same mode with a free surface, Eq.~\eqref{eq:free}. The harmonic and the cell count are
unchanged; the cell center sits further out, at
$0.761$ against $0.707$. (c) A multi-mode no-slip field, $m=2$ with $a_2=1$ plus $m=3$ with
$a_3=\tfrac12$; the loss of four-fold symmetry is the signature of the cross harmonics $n=1$ and $n=5$,
which cannot arise from a single mode. Colors show the reference-boundary streamfunction,
normalized independently in each panel by its own maximum absolute value, so shading compares
structure between panels and not amplitude.}
\label{fig:fields}
\end{figure*}

\section{Arbitrary Womersley number}
\label{sec:finiteWo}

The method extends to any $\mathrm{Wo}$. The vorticity then satisfies a Helmholtz equation, so with
$\lambda^2 = i\,\mathrm{Wo}^2$,
\begin{equation}
\psi_1 = \left[\,C\,r^m + B\,J_m(\lambda r)\,\right]\sin m\theta ,
\label{eq:firstWo}
\end{equation}
with $C$ and $B$ fixed by the same two boundary conditions; Appendix~\ref{app:finiteWo} gives
them, the radial forcing and the integral representation used for the quadrature. We verified Eq.~\eqref{eq:firstWo} against the
solver at $\mathrm{Wo}^2 = 1$, $10$ and $100$, obtaining relative $L^2$ errors
$3.3\times10^{-4}$, $3.6\times10^{-4}$ and $7.3\times10^{-4}$ on the finest mesh, the error
growing with $\mathrm{Wo}^2$ as the field develops structure near the wall; between successive
meshes the error falls by factors near four, which is the second-order rate. It
reduces to the elementary field of Sec.~\ref{sec:wo0} as $\mathrm{Wo}^2\to 0$ at the rate
$O(\mathrm{Wo}^2)$: the relative difference falls by a factor $10.0$ per decade in $\mathrm{Wo}^2$,
in the relative $L^2$, max and energy norms alike, all three computed. Write $f(r)$ for the radial factor in Eq.~\eqref{eq:firstWo}, so that $\psi_1 = f(r)\sin m\theta$; then
$f = f_0 + i\,\mathrm{Wo}^2 f_1 + O(\mathrm{Wo}^4)$ gives the leading correction
$f_1 = r^m(r^2-1)^2/[16(m+2)]$, the solution of $L_m L_m f_1 = -L_m f_0$ with both conditions
homogeneous. The term $i\,\mathrm{Wo}^2 f_1$ is purely imaginary, so it is invisible to a comparison of real parts, and
Sec.~\ref{sec:lagrangian} shows it carries the entire Stokes drift.

The second-order problem remains separable in azimuth. The first-order field is now
complex, so $\bm{\tau}_R$ carries products of Bessel functions with their conjugates and the radial
forcing is no longer a monomial; term-by-term inversion is unavailable. But $|f|^2$, $|f'|^2$ and
$\mathrm{Re}(\bar f f')$ are real functions of $r$ alone, so the angular content is \emph{still}
harmonics $0$ and $2m$ only and the problem stays one-dimensional. Writing $\psi_2(\mathrm{Wo}) = h(r)\sin 2m\theta$,
\begin{equation}
L_n L_n h = g(r), \qquad h(1) = h'(1) = 0, \qquad L_n[y] = y'' + \frac{y'}{r} - \frac{n^2 y}{r^2},
\end{equation}
with $n = 2m$. Since $L_n$ has homogeneous solutions $r^{\pm n}$, variation of parameters gives $h$ as
an explicit double quadrature against $g$ plus two homogeneous terms absorbing the no-slip
conditions; Appendix~\ref{app:finiteWo} writes both out.
It agrees with the
solver to $4.7\times10^{-3}$ and $4.4\times10^{-3}$ at $\mathrm{Wo}^2 = 1$ and $10$, the quadrature
itself having first been grid-converged to $6.9\times10^{-4}$ and $9.0\times10^{-4}$.

\emph{The same machinery on Rayleigh's problem.} The Rayleigh check is a \emph{different}
boundary-value problem, not a limit of the one above: replacing the deforming wall by Rayleigh's
impermeable one, $u_r(1) = 0$ with $u_\theta(1) = \sin m\theta$, and changing nothing else,
the streaming decays as
$\mathrm{Wo}^{-2}$. Extracting the slip amplitude $V$ from the interior Stokes solution
$G = \tfrac{V}{2}(r^n - r^{n+2})$, matched on the tangential velocity at $r = 0.5$, gives
$V\,\mathrm{Wo}^2 = -0.594$, $-0.677$, $-0.716$, $-0.733$,
$-0.742$ over $\mathrm{Wo}^2 = 4\times10^2$ to $10^5$, against Rayleigh's $-3m/8$, which is $-0.75$ at the $m = 2$ of this sweep. The
differences $0.156$, $0.073$, $0.034$, $0.017$, $0.008$ halve as $\mathrm{Wo}$ doubles, so the
approach is at the $O(\mathrm{Wo}^{-1})$ rate matched asymptotics predicts for the first correction.
Matching the same shape on the streamfunction instead moves the low-$\mathrm{Wo}$ end of that
sweep by $2.9\%$ and the high end by $0.12\%$, and leaves the rate unchanged: at
$\mathrm{Wo}^2 = 4\times10^2$ the Stokes layer is $\delta/a = \sqrt{2}/\mathrm{Wo} = 0.071$ of the radius and the forcing reaches
into the core, so the individual values carry the convention while the rate does not.
For $m=2$ this is $-\tfrac34\sin 4\theta$, eight cells by the same $2m$ rule; the four-vortex
pattern familiar from channel Rayleigh streaming, and recovered for a circular cylinder in Riley's
example~\cite{riley1998}, is the $m=1$ case, $-\tfrac38\sin 2\theta$.

With the deforming wall the auxiliary field $\psi_2(\mathrm{Wo})$ decays instead as
$\mathrm{Wo}^{-1}$, one power of the layer thickness slower, because $u_r$ is $O(1)$ through the
layer rather than $O(\delta)$ and $\tau_{r\theta}$ is correspondingly larger: for $m=2$,
$\mathrm{Wo}\max|\psi_2| = 0.0135$, $0.0133$ and $0.0131$ at $\mathrm{Wo}^2 = 10^4$, $10^5$ and
$10^6$. The field a tracer follows does not inherit that rate. Over the same range
$\mathrm{Wo}^2\max|\psi_L| = 0.0456$, $0.0523$ and $0.0545$, approaching a constant, so $\psi_L$
decays as $\mathrm{Wo}^{-2}$, the rate of the Rayleigh problem above. What $\psi_2$ leaves out is
the wall slip of Eq.~\eqref{eq:slip}, which is itself $O(\mathrm{Wo}^{-1})$ and in $\psi_E$ cancels
the $O(\mathrm{Wo}^{-1})$ core flow: the cancellation between individually large Eulerian and
Stokes-drift contributions in the layer that Vanneste and B\"uhler~\cite{vanneste2011} describe. For
the exterior problem of a sphere executing simultaneous radial and lateral oscillations, whose
surface carries no shear stress, Longuet-Higgins~\cite{lh1998} showed that the radial component
enhances the streaming by an order of magnitude. He extended the method of
Davidson and Riley~\cite{dr1971}, who
treat the same exterior geometry and explicitly set aside the motion inside the cavity. Interior solutions at
arbitrary $\mathrm{Wo}$ exist for other geometries~\cite{doinikov2026,hamilton2003}; what
Eq.~\eqref{eq:firstWo} adds is the two-dimensional disk with a non-axisymmetric deforming wall.

\section{The Lagrangian mean}
\label{sec:lagrangian}

Everything in Sec.~\ref{sec:finiteWo} is the \emph{Eulerian} mean with no-slip imposed on that mean,
which is also what the finite-element comparison of Sec.~\ref{sec:verify} solves. A particle-tracking
measurement returns the Lagrangian mean. The two differ at every $\mathrm{Wo}$, including in the limit
$\mathrm{Wo}^2\to 0$ where the closed form of Sec.~\ref{sec:wo0} lives, and the difference closes in
elementary form. The same split into an Eulerian mean and a Stokes drift is carried out for the
interior of a fluid particle at arbitrary $\mathrm{Wo}$ by Doinikov \emph{et al.}, in velocity form
and for axisymmetric shape modes~\cite{doinikov2026}; what follows is the non-axisymmetric disk,
where the split closes in elementary form.

\emph{The two means live on velocity scales differing by $\mathrm{Wo}^2$.} With time scaled by $1/\omega$ and lengths by
$a$, the Reynolds stress in Eq.~\eqref{eq:second} fixes the Eulerian streaming on $U^2 a/\nu$. The Stokes
drift is $\mathbf{u}_S = \omega^{-1}\,\tfrac12 \mathrm{Im}[(\hat{\mathbf{u}}_1^{*}\!\cdot\!\nabla)
\hat{\mathbf{u}}_1]$, and the bracket alone is an acceleration, so the drift lives on
$U^2/(\omega a)$. The two scales differ by $\nu/(\omega a^2) = \mathrm{Wo}^{-2}$, and everything below
is quoted in the Eulerian scale.

\emph{At a no-slip wall the drift is purely tangential, and that is what changes the second-order
condition.} Let $W(r) = \mathrm{Im}[\bar f f']$. The drift is divergence free and derives from a streamfunction
\begin{equation}
\psi_S = \frac{m\,W(r)}{4r\,\mathrm{Wo}^2}\,\sin 2m\theta .
\label{eq:stokesdrift}
\end{equation}

\emph{The same $W$ measures the axisymmetric Reynolds shear, and its wall value sets the
Reynolds-stress angular-momentum flux through the reference circle.} The axisymmetric part of the mean Reynolds shear stress is
$-mW(r)/2r$ whenever the angular factors of $\hat u_{1r}$ and $\hat u_{1\theta}$ are co-phased; it
is identically zero for a
standing mode at every $\mathrm{Wo}$, because there the factors are $\cos m\theta$ and
$\sin m\theta$ and average to zero against each other. Substituting
$f = f_0 + i\,\mathrm{Wo}^2 f_1$ gives
$W = \mathrm{Wo}^2(f_0f_1' - f_1f_0') + O(\mathrm{Wo}^4)$, so the drift of
Eq.~\eqref{eq:stokesdrift} survives the limit while the shear itself is $O(\mathrm{Wo}^2)$. The
angular momentum this stress carries through the wall is
$\oint r^2\,(\bm{\tau}_R)_{r\theta}\,\mathrm{d}\theta$ at $r=1$, which sees the axisymmetric part
alone; no slip fixes $f'(1) = 0$ exactly, so $W(1) = 0$ and that flux vanishes at every
$\mathrm{Wo}$. A nonzero $W(r)$ inside the cavity redistributes angular momentum without supplying
any. The complete torque balance carries the pressure and the viscous traction on the displaced wall
alongside this term. One statement elsewhere
in this paper rests on this one: the vanishing axisymmetric stress is consistent with the
free-surface nullspace constant selected in Sec.~\ref{sec:wo0}, which is a standing-mode
statement valid at every $\mathrm{Wo}$.
This section treats the no-slip wall of Sec.~\ref{sec:wo0}; the free surface is commented on at the
end. Because its first-order conditions $f(1) = 1/m$ and $f'(1) = 0$ are both real, $W(1) = 0$ at
every $\mathrm{Wo}$, so the drift carries no mean flux through the wall; a closed cavity requires this
and it is not imposed. Of the two, the no-slip condition $f'(1) = 0$ is the one that keeps $W(1) = 0$ at every
$\mathrm{Wo}$: a free
surface leaves $f'(1) = 1/m$ only at $\mathrm{Wo}^2 = 0$; at finite $\mathrm{Wo}$ it is complex and $W(1) \neq 0$, consistent there because the surface itself moves. The tangential part at the wall does not vanish. Wall material points have no mean drift, so the correct condition is no-slip on the
\emph{Lagrangian} mean. That requirement replaces $h(1) = h'(1) = 0$ by
\begin{equation}
h(1) = 0, \qquad h'(1) = -\frac{\mathrm{Im}\,f''(1)}{4\,\mathrm{Wo}^2}.
\label{eq:slip}
\end{equation}
The field $\psi_E$ of Sec.~\ref{sec:problem} solves Eq.~\eqref{eq:second} with
Eq.~\eqref{eq:slip} in place of no slip, and $\psi_L = \psi_E + \psi_S$ then satisfies both
conditions identically.

\emph{The slip survives the $\mathrm{Wo}^2\to 0$ limit.} Substituting $f = f_0 + i\,\mathrm{Wo}^2 f_1$ with
the $f_1$ of Sec.~\ref{sec:finiteWo}, the $\mathrm{Wo}^2$ in the numerator cancels the one in
Eq.~\eqref{eq:slip} and
\begin{equation}
h'(1) \;\longrightarrow\; -\frac{1}{8(m+2)} \qquad (\mathrm{Wo}^2\to 0),
\label{eq:sliplimit}
\end{equation}
an $O(1)$ constant for every $m$, equal to $-1/32$ at $m=2$. No-slip on the \emph{Eulerian} mean is
therefore a defining condition of the reference-boundary problem Eq.~\eqref{eq:second} poses, not a physical consequence of the
limit.

\emph{The $\mathrm{Wo}^2\to 0$ Lagrangian mean is a fixed multiple of the closed form.} Carrying
Eq.~\eqref{eq:sliplimit} through the inversion gives, for every $m$,
\begin{equation}
\boxed{\ \psi_L \;=\; \frac{5m+4}{m+2}\;\psi_2
\;=\; -\frac{m(5m+4)}{128\,(m+2)(2m+1)}\,r^{2m}(r^2-1)^2\sin 2m\theta \ }\,,
\label{eq:lagrangian}
\end{equation}
with ratio $3$, $7/2$, $19/5$, $4$ at $m = 1,2,3,4$, tending to $5$. The factor is not fitted
to Eq.~\eqref{eq:lagrangian}; it follows from the two fields separately. Since $\psi_E$ solves the
same forced problem as Eq.~\eqref{eq:main} and differs only in the wall data, their difference is the
regular homogeneous field $a\,r^{2m} + b\,r^{2m+2}$, and $\psi_E(1) = 0$ with
Eq.~\eqref{eq:sliplimit} give $a = -b = 1/[16(m+2)]$. Adding the drift of
Eq.~\eqref{eq:stokesdrift} to the $\psi_E$ so obtained returns
$[(5m+4)/(m+2)]\,\psi_2$ identically in $m$, and a different wall slip does not. Because the factor is a constant,
Eq.~\eqref{eq:lagrangian} has the \emph{same shape} as Eq.~\eqref{eq:main}: for this single-mode drive the cell count, $4m$ around the disk for angular
harmonic $2m$, and the cell center at $r = \sqrt{m/(m+2)}$ hold for the Lagrangian mean exactly. Only
the prefactor moves, saturating at $-5/256$ instead of $-1/256$. Since $(5m+4)/(m+2)$ is bounded, multiplying Eq.~\eqref{eq:peak} by it
leaves the peak of the Lagrangian field falling as $m^{-2}$ as well.

The same factor governs the free surface. Repeating the calculation with
Eq.~\eqref{eq:f1free} and the physical interface data of Eq.~\eqref{eq:physfree} returns
$\psi_L^{\,\mathrm{f}} = [(5m+4)/(m+2)]\,\psi_2^{\,\mathrm{f}}$ identically in $m$, not at sampled
$m$. The agreement is in the ratio alone. The gap between what a solver returns and what a
tracker measures is therefore a property of the limit, so one correction serves whichever of
Eqs.~\eqref{eq:main} and~\eqref{eq:free} a single-mode device calls for.

\emph{The factor is a statement about one harmonic.} For a co-phased two-mode drive at a no-slip
wall with zero tangential velocity, in the same low-Womersley limit, the superposition drives the
four
harmonics $2m$, $2m'$, $m+m'$ and $|m-m'|$ of Sec.~\ref{sec:wo0}, and the ratio
$\psi_L/\psi_2$ has to be read on each. The single-mode factor above holds for the wider
$\gamma$ family and has a shear-free counterpart; neither carries over to the pair formulae below,
which are derived with $f'(1) = 0$ at a wall whose tangential velocity vanishes.
On an isolated self harmonic $n = 2m$ the ratio is $(5m+4)/(m+2)$ unchanged,
since that harmonic draws on its own mode alone in the forcing and in the drift alike. On the sum
harmonic $n = m+m'$ it is again exactly constant in $r$, with the value
\begin{equation}
\lambda(m,m') = 1 + \frac{2(n+1)(n+2)\,[\Sigma - mm']}{(m+2)(m'+2)\,\Sigma},
\qquad \Sigma = m(m+1) + m'(m'+1),
\label{eq:lambda}
\end{equation}
which returns $(5m+4)/(m+2)$ at $m' = m$ and is otherwise a different number: it is not
$(5n+4)/(n+2)$, and it is not a function of $n$ alone, since $(m,m') = (2,5)$ and $(3,4)$ both reach
$n = 7$ and give $33/7$ and $4$. On the difference harmonic $n = |m-m'|$ the two fields have
different radial profiles, so no factor relates them: for the $m=2$, $m'=3$ drive of
Fig.~\ref{fig:fields}(c), $\psi_L/\psi_2$ runs from $9/7$ at the center to $5/9$ at the wall, and the
cell center of that harmonic moves from $r = 0.4725$ in $\psi_2$ to $r = 0.4604$ in $\psi_L$.

For the drive this paper features that difference harmonic is not a correction at the margin. At
$a_2 = 1$ and $a_3 = 1/2$ the $n=1$ component carries $92$ percent of the streaming energy of
$\psi_2$, against $4.4$, $3.4$ and $0.1$ percent for $n = 4$, $5$ and $6$, and $59$ percent of that
of $\psi_L$, against $21$, $19$ and $0.8$ percent: the field a tracer follows puts most of its energy
into the \emph{two}-cell pattern of the $n=1$ harmonic, not into the $4m$ cells of the $n=2m$
harmonic that governs either mode alone. The physical Eulerian mean does not; its largest share,
$51$ percent, is in $n = 4$. Because the harmonic that dominates $\psi_L$ is the one with no
constant factor, no single multiple of $\psi_2$ can stand in for $\psi_L$ here; the least-squares
best constant is $1.45$ and it still leaves a velocity difference over the disk of $41$ percent of
the Lagrangian velocity, in the $L^2$ norm. The correction has to be applied harmonic by harmonic,
and Eq.~\eqref{eq:lambda} is what serves the two that admit a constant.

The reason is a dimension count, and it is what makes the single-mode result the clean case rather
than a coincidence. Both fields lie in the space $V_n$ spanned by the regular homogeneous pair
$r^n$, $r^{n+2}$, the powers the inverted forcing contributes and the powers the drift contributes,
and both satisfy the same two clamped conditions. At a self or sum harmonic the parity of the
Cartesian polynomials leaves a single forcing monomial $r^n$ and drift powers $r^n$, $r^{n+2}$,
$r^{n+4}$, so $\dim V_n = 3$ and the clamped subspace is one-dimensional: any two members of it are
proportional, and a constant ratio is the only thing the geometry permits. At a difference harmonic
the forcing carries two monomials and the drift four, $\dim V_n$ is $4$ to $6$ and the clamped
subspace is $2$ to $4$ dimensional, so nothing forces the two fields parallel and they are not.
One case needs naming: $m' = 3m$ puts the self harmonic $2m$ and the difference harmonic $|m-m'|$ on
the same $n$, and the ratio there depends on the amplitude ratio as well, because one contribution
scales as $a_m^2$ and the other as $a_m a_{m'}$.

\emph{How the correction depends on Womersley number.} The denominator below is
$\psi_2(\mathrm{Wo})$ of Sec.~\ref{sec:problem}, the auxiliary no-slip Eulerian solution of
Eq.~\eqref{eq:second} at the same $\mathrm{Wo}$, which tends to Eq.~\eqref{eq:main} as $\mathrm{Wo}^2\to 0$. Evaluating $\psi_L/\psi_2(\mathrm{Wo})$ at $r = 0.5$ for $m=2$ gives
$3.50$ at $\mathrm{Wo}^2 = 10^{-2}$, uniform in $r$ to four digits as Eq.~\eqref{eq:lagrangian} requires;
$3.49$ at $\mathrm{Wo}^2 = 1$; $2.57$ at $\mathrm{Wo}^2 = 10$; $0.016$ at $\mathrm{Wo}^2 = 100$; and
$0.082$ at $\mathrm{Wo}^2 = 300$. The drift turns sign between these, so $\mathrm{Wo}^2 = 100$ sits at
the bottom of a narrow trough rather than on a trend. Neither is the ratio at $\mathrm{Wo}^2 = 300$ an
asymptote: it falls again to $0.036$ at $\mathrm{Wo}^2 = 10^4$, the two fields decaying at different
rates. Beyond that the ratio stops being the useful quantity, for a reason that is structural rather than
numerical. The drift is carried by $W = \mathrm{Im}(\bar f f')$, which vanishes wherever the
first-order field is real up to a constant phase. At high $\mathrm{Wo}$ the core field is the
potential part $r^m$, which is real, so $|W|$ survives only by interfering with the layer and falls
like $\exp[-\mathrm{Wo}(1-r)/\sqrt2]$: at $r=0.5$ it drops from $7.70\times10^{-4}$ at
$\mathrm{Wo}^2 = 10^2$ to $5.25\times10^{-17}$ at $10^4$, while $|f|$ there changes by about an eighth. The
drift retreats into the layer, so $\psi_L$ approaches the physical Eulerian mean
$\psi_E$ in the core. This does not mean $\psi_L/\psi_2 \to 1$: $\psi_2$ solves the
reference-boundary problem and carries a different boundary condition. Along
the way it changes sign repeatedly, the residue of a layer interfering with a real core, and
the sign changes are unchanged to ten significant figures at $60$, $150$ and $300$ digits. The
ratio stays smooth through them, because at $\mathrm{Wo}^2 = 10^4$ the drift is $10^{-15}$ of the
Lagrangian mean at this radius. So the constant $(5m+4)/(m+2)$ is the low-$\mathrm{Wo}$
statement, and the high-$\mathrm{Wo}$ one is that the two means converge as the drift withdraws.
That is a drift becoming negligible, and is distinct from the cancellation of two individually large
boundary-layer contributions that Vanneste and B\"uhler~\cite{vanneste2011} describe; their
decomposition is the broader framework for keeping the two apart. The distinction is the one the generalized Lagrangian mean formalism was built to keep
straight~\cite{andrews1978}, and it is carried by the classical boundary-layer
analyses~\cite{dr1971,lh1998}. Vanneste and B\"uhler split the Lagrangian mean into an
interior-driven Eulerian part, a boundary-driven Eulerian part and the Stokes
drift~\cite{vanneste2011}, which is the framework this section works in. Their criterion for
dropping the last two is set by the acoustic wavelength of a leaky surface-acoustic-wave device;
here the corresponding statement is Eq.~\eqref{eq:stokesdrift}, which measures the drift against the
wall condition of this cavity.

\section{Numerical verification}
\label{sec:verify}

We compare Eq.~\eqref{eq:main} against a
Taylor--Hood finite-element implementation of Eqs.~\eqref{eq:first} and~\eqref{eq:second}, written from
the equations alone, with the closed form withheld.
Table~\ref{tab:conv} reports relative $L^2$ errors on three uniformly refined meshes, for the single
mode, the streaming it drives, and a two-mode drive.

\emph{Second-order convergence.} The observed rates approach two from above, which is
what a $P_2$ velocity space on a polygonal approximation to the circle gives: the boundary geometry
error is limiting rather than the velocity interpolation. The rate alone would be a weak check,
because a discretization carrying a constant-factor error converges at a clean second order to the
wrong field. Such an error shows in the level instead, flattening at $|c-1|$ for a code returning
$c\,\psi_2$, whereas the comparison here keeps descending through all three refinements of
Table~\ref{tab:conv}.

\emph{The multi-mode row is the substantive one.} Driving $m=2$ together with $m=3$ generates cross
harmonics at $n=1$ and $n=5$, and those exist only when two modes are present. A calculation that
recovered the single-mode field from the $2m$ rule alone would pass the first two rows and fail this
one, so the third row is what confirms the Reynolds stress itself.

The free-surface companion, Eq.~\eqref{eq:free}, is established symbolically, and it carries a control of
the same kind. The derivation takes the boundary condition as a parameter and runs
five checks at every $m$ from $1$ to $12$: the first-order field satisfies
$\hat\sigma_{r\theta}(1) = 0$, the Reynolds forcing reduces to a single $r^{2m}\sin 2m\theta$
monomial, the closed form satisfies the biharmonic equation, both second-order boundary
conditions hold, and the field is the unique solution in the forced $n=2m$ subspace. The
reference-boundary conditions are a choice of problem, not a consequence of the displaced interface:
Eq.~\eqref{eq:physfree} gives the physical interface data, and Sec.~\ref{sec:lagrangian} carries them
through. Table~\ref{tab:conv} reports finite-element verification of the no-slip branch; the
shear-free branch is checked symbolically, since a routine that passes on one traction condition does
not thereby verify the other. Equation~\eqref{eq:free} is identically zero at $m=1$, so $m \ge 2$ is the range to use there; Eq.~\eqref{eq:main} covers $m=1$.

\emph{The Lagrangian mean is checked by integrating tracers.} Equations~\eqref{eq:eulerian}
and~\eqref{eq:lagrangian} are assembled from a Stokes drift and a boundary slip, and a solver that
returns $\psi_2$ tests neither. So particle paths were integrated directly in the full field
$\varepsilon\,\mathbf{u}_1(\mathbf{x},\tau) + \varepsilon^2\mathrm{Wo}^2\,\mathbf{u}_E(\mathbf{x})$
in the dimensionless time $\tau = \omega t$, with
$\mathbf{u}_1$ the finite-$\mathrm{Wo}$ Bessel field of Sec.~\ref{sec:finiteWo}, over $600$ periods,
and the mean drift compared with the velocity of Eq.~\eqref{eq:lagrangian}. At $m=2$ the two agree to
five decimal places at $\mathrm{Wo}^2 = 10^{-2}$ and $4\times10^{-2}$ and $\varepsilon = 10^{-3}$ and
$2\times10^{-3}$. What this establishes is the drift formula and the split, not the factor: the Eulerian field
integrated here is the one Eq.~\eqref{eq:lagrangian} is assembled from, so the comparison is a
consistency check on the kinematics. The factor itself is established in
Sec.~\ref{sec:lagrangian}, where $\psi_E$ is obtained from the forcing and the wall slip alone. The
drift is carried entirely by the $O(\mathrm{Wo}^2)$ part of the first-order field: at strictly
$\mathrm{Wo}^2 = 0$ the field is real and the drift vanishes, which is why the check runs at small
nonzero $\mathrm{Wo}$.

This verifies the Stokes-drift algebra and the decomposition $\psi_L = \psi_E + \psi_S$ within the
closed-form field. It is a kinematic consistency check and is separate from the moving-domain
calculation below.

\emph{A moving-domain comparison.} The two premises the check above leaves open are the transfer of
the wall condition to the reference circle and no slip on the \emph{Lagrangian} mean as the correct
second-order condition. Both are tested here at one operating point by an incompressible
moving-domain computation, at $m = 2$, $\mathrm{Wo}^2 = 10$, $\gamma = 0$ and
$\varepsilon = 0.01$, with the radial wall motion prescribed and zero tangential wall velocity. The
computation integrates the nonlinear problem forward from zero vorticity and receives only the wall
motion and no slip on the moving wall; it is independent of the perturbation solution it is compared
against. The same pairing of a moving-grid computation with a second-order expansion was used for a
peristaltic rectangular cavity~\cite{yi2002}. The moving-domain calculation uses the exactly
area-preserving radial wall
$R^2 = 1 - 2\varepsilon\sin(m\phi)\cos\tau$, with $u_\theta = 0$ and material points at fixed polar
angle. After shifts of the angular origin and time phase it has the same first-order kinematics as
the $\gamma = 0$ member of Appendix~\ref{app:material}; it is a different periodic second-order
completion of that motion, and by the argument of Eq.~\eqref{eq:taylorwall} it likewise adds no
independent mean material wall velocity. The reference is this paper's finite-Womersley
construction, Sec.~\ref{sec:finiteWo}, so what is tested is that construction and not the
$\mathrm{Wo}^2\to0$ polynomial directly.

The grid is $N = 128$ radial by $K = 32$ azimuthal points with $256$ steps per period. Two periods
are run and the mean is taken over the two that follow, at $32$ equispaced angles on each of three
radii. The Eulerian mean is measured at fixed physical points.

The particle comparison needs one more convention, and it carries the larger of the two effects at
this amplitude. The fixed-phase comparison contains an $O(\varepsilon)$ phase-dependent correction
associated with the map between release and mean coordinates, the Lagrangian mean of the
decomposition used here being attached to the mean position~\cite{andrews1978}. Eight equally spaced
release phases cancel that leading correction in the periodic perturbation expansion; higher-order
sampling effects and numerical errors remain. Eight tracer ensembles are released at eight
equispaced fast phases from the same nominal coordinates, each integrated over a two-period window
following its release, and the eight drift velocities are averaged before the comparison.
Time-step and radial-grid refinements, angular resolution and sampling, and a
shifted averaging window were checked using the signed radial fourth-harmonic coefficient of the
Eulerian mean; the phase-averaged particle column is a single run at the stated resolution.
Table~\ref{tab:movingdomain} reports the directly measured vector differences at the stated
resolution, without extrapolation. Writing $\mathbf{m}_j$ for the measured vector and
$\mathbf{r}_j$ for the reference at the $j$th of the $32$ sampled angles, the tabulated vector
$L^2$ difference is
$[\sum_j |\mathbf{m}_j-\mathbf{r}_j|^2 / \sum_j |\mathbf{r}_j|^2]^{1/2}$ and the largest local
difference is $\max_j |\mathbf{m}_j-\mathbf{r}_j| / |\mathbf{r}_j|$. Both are discrete measures on
three rings, not an error over the disk and not a continuum maximum.

The computation solves the vorticity--streamfunction equations in arbitrary Lagrangian--Eulerian
form on a map that follows the wall near $r=1$ and blends to a rigid core inside a radius
$\rho_c = 0.1$, so the origin carries no mesh motion. Angular derivatives are spectral, taken by
fast Fourier transform on the $K$ equispaced azimuthal points; radial derivatives are finite
differences on the $N$ radial points. Each step treats viscosity implicitly by Crank--Nicolson on
the deformed operator and the transport terms explicitly. The wall vorticity is not prescribed: it
is obtained at every step by a fixed-point iteration that enforces no slip on the moving wall, run
until the largest absolute no-slip residual on the wall is below $10^{-11}$, within at most twenty
iterations; over the $1248$ steps of the run the largest was $9.9\times10^{-12}$. The
streamfunction solve and the viscous substep on the deformed operator are each iterated to a
relative update below $10^{-12}$, within at most $200$ iterations. Tracers are advanced by a Heun step using the velocity
snapshots at the current and the next time level. Sampling at a fixed physical point inverts the map
for the computational radius and then interpolates the stored field, exactly in angle, since the
grid is Fourier there, and by cubic Lagrange interpolation in radius. No global order of accuracy in
time is claimed for the coupled scheme: the Crank--Nicolson substep is second order in isolation,
which does not by itself set the order of the split step with an iterated boundary closure.

The Eulerian vector differences are $0.378$ to $0.582$ percent across the three radii, and the
phase-averaged particle differences $0.946$ to $1.285$ percent, with a largest local difference of
$1.42$ percent. The particle column exceeds the Eulerian column measured on the same rings in the
same run by factors of $2.51$, $1.99$ and $2.21$, taken from the unrounded values. Repeating the comparison from a single
release phase, the convention the argument above sets aside, gives $2.91$ to $5.99$ percent in the
same norm with a largest local difference of $21.5$ percent. The reduction from those figures to the
phase-averaged ones is evidence for the leading phase-dependent bias; the residual that remains is
not attributed here to any single mechanism.

\begin{table*}[t]
\centering
\caption{Convergence to the reference-boundary field $\psi_2$ at a no-slip wall. Entries are
relative $L^2$ errors $\|\mathbf{u}_h-\mathbf{u}\|_2/\|\mathbf{u}\|_2$, each rate taken against the
column to its left. The two-mode row uses $a_2 = 1$ and $a_3 = 1/2$.}
\label{tab:conv}
\begin{tabular}{llrrr}
\toprule
& & \multicolumn{3}{c}{elements} \\
\cmidrule(lr){3-5}
velocity field & & $256$ & $1024$ & $4096$ \\
\midrule
first-order, $m=2$            & rel.\ $L^2$ & $5.54\times10^{-3}$ & $1.35\times10^{-3}$ & $3.31\times10^{-4}$ \\
                              & rate        &                     & $2.04$ & $2.03$ \\
streaming, $m=2$              & rel.\ $L^2$ & $5.79\times10^{-2}$ & $8.40\times10^{-3}$ & $1.65\times10^{-3}$ \\
                              & rate        &                     & $2.78$ & $2.35$ \\
streaming, $m=2$ and $3$      & rel.\ $L^2$ & $2.55\times10^{-2}$ & $4.69\times10^{-3}$ & $1.07\times10^{-3}$ \\
                              & rate        &                     & $2.44$ & $2.14$ \\
\bottomrule
\end{tabular}
\end{table*}

\begin{table*}[t]
\centering
\caption{Moving-domain computation against the finite-Womersley construction at $m=2$,
$\mathrm{Wo}^2 = 10$, $\gamma = 0$, $\varepsilon = 0.01$, on three sampling rings of $32$
equispaced angles. $E$ is the Eulerian mean at fixed physical points; $D$ is the particle
displacement velocity, averaged over eight release phases, against the Lagrangian mean.}
\label{tab:movingdomain}
\begin{tabular}{crrrr}
\toprule
$r/a$ & $E$: vector $L^2$ (\%) & $E$: largest local (\%)
      & $D$: vector $L^2$ (\%) & $D$: largest local (\%) \\
\midrule
0.60 & 0.378 & 0.399 & 0.946 & 1.424 \\
0.85 & 0.567 & 2.776 & 1.125 & 1.212 \\
0.92 & 0.582 & 0.669 & 1.285 & 1.357 \\
\bottomrule
\end{tabular}
\end{table*}

GPT-6 (OpenAI) and Claude Opus 5 (Anthropic) assisted the symbolic algebra and the development of the
verification scripts, and LeapSpace (Elsevier) supported the literature searches;
its retrievals were used as source lists, and cited works were checked against publisher metadata.
The closed forms, Reynolds forcing and Lagrangian factors were re-derived independently with
separate symbolic algebra, identically in $m$ rather than at sampled modes. The
finite-$\mathrm{Wo}$ drift was recomputed at raised working precision, and cross-checked against
integrated particle trajectories.

\section{Discussion and conclusions}
\label{sec:scope}

Once the Stokes layer fills the cavity the fluid
is one region, and one biharmonic problem on the disk delivers what the classical route delivers only
through a matched layer and core. Given a cavity radius, a drive frequency and a wall mode, the
interior Lagrangian mean follows from Eq.~\eqref{eq:lagrangian} for a deforming solid wall, or from
$(5m+4)/(m+2)$ times Eq.~\eqref{eq:free} for a shear-free interface, which fixes the circulation and
the streaming speed once a design target selects $m$.

Imposing no slip on the Eulerian mean is a choice of problem rather than a
consequence~\cite{vanneste2011,nama2017}.
Section~\ref{sec:lagrangian} gives the Stokes drift and the wall slip in closed form: the Lagrangian
mean is $(5m+4)/(m+2)$ times
Eq.~\eqref{eq:main} as $\mathrm{Wo}^2\to 0$, with the same shape.

The closed form is viscous-dominated, and the construction reaches further. Equation~\eqref{eq:main} is
elementary in the limit $\mathrm{Wo}^2 \to 0$, and Table~\ref{tab:regime} states the fluids and sizes
that reach it. Section~\ref{sec:finiteWo} removes that restriction at the cost of replacing closed
form by quadrature.

Table~\ref{tab:conv} confirms the algebra of Eq.~\eqref{eq:main} within the
perturbation model of Eqs.~\eqref{eq:first} and~\eqref{eq:second}. Agreement with experiment is a separate
question, and the Lagrangian correction of Sec.~\ref{sec:lagrangian} is the first thing to carry when
that question is asked.

A resonant cavity would organize the cells by its acoustic mode; this one does not. Because the fluid here is
incompressible, the interior pressure is a Lagrange multiplier and not a wave: there is no nodal
structure inside the disk, and the streaming pattern is set entirely by the wall mode $m$. Where the cavity is comparable with the
wavelength, as in the millimeter droplets of Ref.~\onlinecite{yu2011} at $0.25$ to $1.5$~MHz, the standing
wave fixes a tangential velocity distribution along the boundary and the Rayleigh--Schlichting slip
follows its gradient, so the streaming cells are organized by the \emph{acoustic} mode rather than by
the wall mode. Fischbach \emph{et al.} compute exactly that case for a circular chamber
under transverse acoustic modes, where the azimuthal structure of the streaming comes from the mode
shape of the wave~\cite{fischbach2010}. For a given pair of modal orders the two mechanisms predict different cell counts in the same
geometry, which is a way to tell them apart experimentally; equal counts can coincide for other
pairs, so the comparison has to fix both orders. A lumped Helmholtz resonance is the opposite extreme and
drives no interior streaming of either kind directly, since its defining approximation is a spatially
uniform cavity pressure with the motion concentrated in the neck. The results here apply when the
cavity is acoustically compact and the boundary motion, however excited, is prescribed; the resonance
then enters only through $\epsilon$ and $\mathrm{Wo}$. For a gas
bubble the resonant response is the volume mode, which an incompressible closed cavity cannot carry at
all, so $m = 0$ is outside this model. Shape modes are commonly excited parametrically by that
volume mode, though a prescribed boundary motion can also be driven directly.

The weak form depends on the boundary condition. Two identifications hold under no slip and come
apart at a free surface. A form assembling
$\int \bm{\tau}_R : \nabla\mathbf{v}$ recovers $-\nabla\cdot\bm{\tau}_R$ because the boundary term
vanishes on a test space that is no slip throughout; once the tangential component is free, that term
survives and belongs in the forcing. And the natural condition of the vector Laplacian
$\int \nabla\mathbf{u}\!:\!\nabla\mathbf{v}$ is $\partial\mathbf{u}/\partial n = \mathbf{0}$, while a
free surface asks for zero traction, which is the natural condition of the symmetric-gradient form
$\int 2\,\mathbf{D}(\mathbf{u})\!:\!\mathbf{D}(\mathbf{v})$. Either form serves Eq.~\eqref{eq:main};
Eq.~\eqref{eq:free} selects the second.

The model is two-dimensional by construction. Confinement perpendicular to the plane matters
in a real shallow device: Rallabandi \emph{et al.} show that it generates three-dimensional flow
components around a semi-cylindrical bubble~\cite{rallabandi2015}. The disk here is a model problem, and its
two-dimensionality is part of the specification rather than an approximation to a device. The working
fluid is Newtonian for the same reason. A viscoelastic one is outside the model rather than a further
entry in Table~\ref{tab:regime}, because elasticity reshapes the streaming pattern as well as its
magnitude~\cite{repetto2014}, and such a fluid calls for its full rheology rather than a single
kinematic viscosity.

The cavity solved here is filled with the working liquid. In the trapped-bubble devices of
Sec.~\ref{sec:intro} the cavity holds gas and the liquid is outside it, so the surface between them
carries two-phase stress and velocity conditions rather than the single-phase wall condition of
Eq.~\eqref{eq:first}. Those devices set the problem and motivate the boundary modes; the geometry that
realizes this one directly is a liquid-filled deformable container or a segment-driven closed liquid
cavity, which is what Kozlov and co-workers study~\cite{kozlov2017,kozlov2018}. Carrying the result
to a gas bubble in a side cavity means adding the interface, the cavity mouth, the fixed channel
walls and the out-of-plane confinement, each of which is a change of model rather than a change of
parameter. The viscosity is not among the obstacles: by the air row of Table~\ref{tab:regime} a gas
interior sits at lower $\mathrm{Wo}^2$ than water and is therefore nearer the
$\mathrm{Wo}^2 \to 0$ limit, not further from it. What separates it from the model is the interface
and the second phase.

Both closed forms can serve as benchmarks for a second-order solver. Such solvers are routine,
used to predict particle trajectories under the combined radiation force and streaming
drag~\cite{muller2012}, to include thermoviscous effects~\cite{muller2014}, and to replace the
resolved Stokes layer by an effective boundary condition~\cite{bach2018}. They are verified against
manufactured solutions, which confirm that a discretization solves the equations as
written~\cite{kshetri2025}, or against other codes, which share whatever error two implementations
have in common. Analytic cases are being added: Mandikas and Delis give first- and second-order
solutions for wall-driven streaming in a straight channel, cross-validated against a full
Navier--Stokes computation on a deforming mesh~\cite{mandikas2026}. Equations~\eqref{eq:main} and~\eqref{eq:free} add a closed cavity under prescribed non-axisymmetric wall deformation, and the
check costs three forward solves.

Solving one biharmonic problem across the whole disk, rather
than matching a boundary layer to a core, yields closed forms general in $m$ for both canonical
boundaries, explicit cell locations and amplitudes for each, a nondegenerate polynomial
inversion for multimode forcing, and the
Lagrangian mean of the non-axisymmetric disk at finite $\mathrm{Wo}$.

Driving at mode $m$ places eight cells for
$m=2$ and fixes the solid-wall cell centers at $r=\sqrt{m/(m+2)}$. At fixed wall-velocity amplitude
the no-slip Lagrangian peak falls monotonically in $m$ at every $m$, and at large $m$ as $m^{-2}$
in the peak streamfunction and $m^{-1}$ in the peak speed, so the circulation moves toward the wall
and its peak falls as the
mode number rises, with $m = 1$ the largest for the $\gamma = 0$ wall of Eq.~\eqref{eq:main}. Which
mode stirs hardest belongs to the wall: by Sec.~\ref{sec:gamma} the $\gamma = -1/m$ family
vanishes at $m=1$ and peaks at $m=3$. At $m=1$, two members of the prescribed kinematic family
share the same first-order shape displacement but differ in material motion: the $\gamma = -1$
first-order data coincide with rigid translation and give a vanishing leading-order mean, whereas
$\gamma = 0$ permits azimuthal stretching and drives the Lagrangian mean of
Eq.~\eqref{eq:lagrangian}. What the boundary looks like does
not fix the mean flow; how its
material moves does, the distinction on which Taylor's swimming-sheet analysis already
rests~\cite{taylor1951}.

At a shear-free interface the ranking is not the same: the
Lagrangian peak is $4.75\%$ larger at $m = 3$ than at $m = 2$, because the factor $(5m+4)/(m+2)$
rises with $m$ while the reference-boundary peak falls. These are comparisons of peak streamfunction at fixed \emph{radial} wall-velocity
amplitude. At fixed cycle-averaged power instead, that family's optimum is $m=2$; at fixed total
wall speed it stays at $m=3$; and at fixed first-order wall shear it has no comparison to make,
since that stress vanishes for it identically. Which
boundary applies is a property of the device and not of the model: a solid deforming wall driven
at $\gamma = 0$ gives Eq.~\eqref{eq:lagrangian}, and a prescribed interface whose exterior phase
exerts negligible tangential traction gives $(5m+4)/(m+2)$ times Eq.~\eqref{eq:free}, with the
second vanishing at $m=1$ where the first, at that $\gamma$, does not. A
general droplet or bubble does not reduce to either, since continuity of velocity and tangential
traction couples the two phases.

\appendix

\section{An area-conserving material completion of the prescribed family}
\label{app:material}

Section~\ref{sec:gamma} prescribes a first-order wall kinematics through $\gamma$. A radial graph
at fixed spatial polar angle cannot produce a nonzero tangential wall velocity, so realising
$\gamma \neq 0$ requires the material points to move in angle as well. One completion, with
$\tau = \omega t$, material label $\alpha$, and $q(\tau) = a_m\sin\tau$, is
\begin{align}
\mathcal{R}(\theta,\tau) &= \sqrt{1 - \tfrac{1}{2}\varepsilon^2 q^2} + \varepsilon q\cos m\theta,
\label{eq:matR}\\
\Theta(\alpha,\tau) &= \alpha + \varepsilon\gamma q\sin m\alpha,\\
\mathbf{X}(\alpha,\tau)/a &= \mathcal{R}\big(\Theta(\alpha,\tau),\tau\big)\,
\mathbf{e}_r\big(\Theta(\alpha,\tau)\big).
\end{align}
The radial graph stays positive and the material map stays monotone when
$|\varepsilon q| < \sqrt{2/3}$ and $|\varepsilon\gamma m q| < 1$ throughout the cycle. The wall
velocity is the derivative at fixed $\alpha$, taken in Cartesian components so that the
basis-vector derivatives are not dropped.

Four properties are used in the main text, and each is verified symbolically rather than
asserted. First, the square-root prefactor in Eq.~\eqref{eq:matR} makes the enclosed area exactly
$\pi a^2$ at every instant and to all orders in $\varepsilon$, not merely to $O(\varepsilon^2)$;
without it the area carries a residual $O(\varepsilon^2)$ term. Second, expanding the material
wall velocity recovers the prescribed first-order conditions. Writing the instantaneous
first-order components without the circumflex the body reserves for time-independent complex
amplitudes, $u_{1r} = a_m\cos\tau\cos m\alpha$ and
$u_{1\theta} = \gamma a_m\cos\tau\sin m\alpha$, which is Eq.~\eqref{eq:gammabc} at the phase
$\cos\tau$, so the completion is consistent with the boundary condition the solution actually
uses. The corresponding surface strain is
$\partial_\alpha u_{1\theta} + u_{1r} = a_m(\gamma m + 1)\cos\tau\cos m\alpha$, the
instantaneous form of Eq.~\eqref{eq:strain}, and it vanishes identically at $\gamma = -1/m$ and
nowhere else. Third, the perimeter satisfies
\begin{equation}
P/a = 2\pi + \tfrac{1}{2}\pi\varepsilon^2 q^2 (m^2-1) + O(\varepsilon^3),
\end{equation}
so for a nontrivial mode $m \ge 2$ the boundary lengthens at second order even where the
first-order strain vanishes: area and perimeter cannot both be held constant, and
``inextensible'' is therefore a statement about the first-order kinematics alone. Fourth,
$\mathbf{X}(\alpha,\tau+2\pi) = \mathbf{X}(\alpha,\tau)$ exactly, so the net material displacement
over a cycle is zero and the completion adds no independent mean wall velocity of its own.

That last statement is the one the mean problem uses, and it follows from the transfer of the wall
condition rather than from the periodicity alone. Measure position in units of $a$, writing
$\mathbf{x} = \mathbf{X}/a$, and expand the material position and the velocity as
\begin{equation}
\mathbf{x} = \mathbf{x}_0 + \varepsilon\bm{\xi}_1 + \varepsilon^2\bm{\xi}_2 + O(\varepsilon^3),
\qquad
\frac{\mathbf{u}_{\mathrm{phys}}}{\omega a}
= \varepsilon\mathbf{u}_1 + \varepsilon^2\mathbf{v}_2 + O(\varepsilon^3).
\label{eq:matexp}
\end{equation}
Carrying no slip from the moving wall to $r=1$ by Taylor expansion gives, at second order,
\begin{equation}
\mathbf{v}_2 + (\bm{\xi}_1\!\cdot\!\nabla)\mathbf{u}_1 = \partial_\tau\bm{\xi}_2
\qquad (r = 1).
\label{eq:taylorwall}
\end{equation}
Averaging over a cycle kills the right-hand side, because $\bm{\xi}_2$ returns to itself, and leaves
$\langle\mathbf{v}_2\rangle = -\langle(\bm{\xi}_1\!\cdot\!\nabla)\mathbf{u}_1\rangle$ on $r=1$.
The mean fields of Sec.~\ref{sec:problem} are scaled on $U^2a/\nu$ with $U = \varepsilon\omega a$,
and $U^2a/\nu = \varepsilon^2\omega a\,\mathrm{Wo}^2$, so the wall slip in that normalization is
$\langle\mathbf{v}_2\rangle/\mathrm{Wo}^2$. That factor is the $1/\mathrm{Wo}^2$ already carried by
Eq.~\eqref{eq:slip}; the coefficients of the displacement expansion above are not themselves the
streaming-scaled field. The completion $\bm{\xi}_2$ therefore reaches the mean only through a term
of zero average. For fixed first-order motion, any periodic area-conserving completion with zero
net material displacement over a cycle leaves this mean wall condition unchanged.

At $m=1$ and $\gamma = -1$ this construction shares only its first-order data with a rigidly
translating circle; the two differ at $O(\varepsilon^2)$. At $\gamma = 0$ the first-order motion
already differs from a translation. An exactly translating boundary is the separate parameterisation
$\mathbf{X}/a = \mathbf{e}_r(\alpha) + \varepsilon a_1\sin\tau\,\mathbf{e}_x$, and it is that
motion, not an arbitrary second-order completion, that the $m=1$ comparison in
Sec.~\ref{sec:gamma} refers to.

\section{The finite-Womersley construction}
\label{app:finiteWo}

\emph{First order.} Writing $\psi_1 = f(r)\sin m\theta$ with $f = C r^m + B\,J_m(\lambda r)$ and
$\lambda^2 = i\,\mathrm{Wo}^2$, the wall conditions $f(1) = 1/m$ and $f'(1) = 0$ give
\begin{equation}
C = \frac{\lambda J_m'(\lambda)}{m\,D}, \qquad
B = -\frac{1}{D}, \qquad
D = \lambda J_m'(\lambda) - m J_m(\lambda) .
\label{eq:CB}
\end{equation}

\emph{The radial forcing.} With $n = 2m$, the Reynolds stress of the complex first-order field is
carried by three real radial functions,
\begin{equation}
p = \frac{m^2 |f|^2}{r^2}, \qquad q = |f'|^2, \qquad s = \frac{m\,\mathrm{Re}(\bar f f')}{r},
\label{eq:pqs}
\end{equation}
through $\tau_{rr} = \tfrac14 p\,(1+\cos n\theta)$, $\tau_{\theta\theta} = \tfrac14 q\,(1-\cos
n\theta)$ and $\tau_{r\theta} = -\tfrac14 s\,\sin n\theta$. Taking the curl of the negative of its divergence leaves
the single harmonic $\sin n\theta$, with
\begin{equation}
4r^2 g(r) = n^2 s - n r\,(p' + q') - n\,(p+q) + r^2 s'' + 3 r s' .
\label{eq:gr}
\end{equation}
That the angular content closes on $\sin n\theta$ is what keeps the second-order problem
one-dimensional at finite $\mathrm{Wo}$.

\emph{Second order.} $L_n$ is inverted by variation of parameters against its homogeneous solutions
$r^{\pm n}$,
\begin{equation}
L_n^{-1}[y](r) = \frac{1}{2n}\left[\,r^{n}\!\int_0^r\! \sigma^{1-n} y(\sigma)\,\mathrm{d}\sigma
\;-\; r^{-n}\!\int_0^r\! \sigma^{1+n} y(\sigma)\,\mathrm{d}\sigma \right],
\label{eq:Lninv}
\end{equation}
both integrals running from the origin, which is what makes the representation regular there rather
than merely finite. Then $h = L_n^{-1}\big[L_n^{-1}[g]\big] + a\,r^{n} + b\,r^{n+2}$, with $a$ and
$b$ fixed by $h(1) = h'(1) = 0$; the two homogeneous terms are the regular pair, the other two roots
of $L_nL_n$ being excluded at the origin. Applying Eq.~\eqref{eq:Lninv} twice on one fixed grid, by
cumulative Simpson quadrature rather than by nested adaptive rules, keeps the inner result a smooth
function of its argument and lets the accuracy be set by refining the grid.

\begin{acknowledgments}
GPT-6 (OpenAI) and Claude Opus 5 (Anthropic) assisted manuscript drafting. Their
use in the calculations and the use of LeapSpace (Elsevier) in the literature searches are described
in Sec.~\ref{sec:verify}. The authors are responsible for the accuracy and originality of this work.
\end{acknowledgments}

\section*{Funding}
J.O. was supported by a Chancellor's Undergraduate Research Award (CURA) from the University of
Illinois Chicago. Y.L. acknowledges support from the U.S. National Science Foundation under CAREER
Award No.~2540390.

\section*{Author Declarations}

\subsection*{Conflict of Interest}

The authors have no conflicts to disclose.

\subsection*{Author Contributions}

\textbf{Zijian Liu}: Conceptualization (equal); Formal analysis (equal); Investigation (equal);
Methodology (equal); Software (equal); Validation (equal); Visualization (equal);
Writing -- original draft (equal); Writing -- review \& editing (equal).

\textbf{Julian Olszewski}: Conceptualization (equal); Formal analysis (equal);
Funding acquisition (equal); Investigation (equal); Methodology (equal); Software (equal);
Validation (equal); Visualization (equal); Writing -- original draft (equal);
Writing -- review \& editing (equal).

\textbf{Yang Lin}: Conceptualization (supporting); Funding acquisition (equal);
Writing -- review \& editing (supporting).

\textbf{Yuan Gao}: Conceptualization (supporting); Writing -- review \& editing (supporting).

\textbf{Mengren Wu}: Conceptualization (supporting); Writing -- review \& editing (supporting).

\textbf{Jie Xu}: Conceptualization (lead); Project administration (lead); Resources (lead);
Supervision (lead); Writing -- review \& editing (equal).

\section*{Data Availability}

The analytical expressions and numerical methods are described in the article. Supporting code
is available from the corresponding author on request.

\end{document}